\documentclass[useAMS,usenatbib]{mn2e}

\pdfoutput=1

\usepackage{ifpdf}
\usepackage{amssymb,amsmath}
\usepackage{graphics}
\usepackage{subfigure}
\usepackage[dvips]{graphicx}
\usepackage{epstopdf}
\usepackage{textcomp}
\usepackage[draft]{hyperref}
\usepackage[english]{babel}
\usepackage{color}
\usepackage{float}
\usepackage{verbatim}
\usepackage[normalem]{ulem}
\usepackage[T1]{fontenc}  
\usepackage{aecompl}

\usepackage{algpseudocode}

\newcommand {\e}{{\rm e}}

\title[Dust sputtering in SN shocks]
{Thermal and nonthermal dust sputtering in hydrodynamical simulations of the multiphase interstellar medium}

\author[C.-Y. Hu et al.]
{
	Chia-Yu Hu$^{1}$\thanks{chu@flatironinstitute.org}, Svitlana Zhukovska$^{2,4}$, 
	Rachel S. Somerville$^{1,3}$, Thorsten Naab$^{2}$\\
	$^1$Center for Computational Astrophysics, Flatiron Institute, 162 5th Ave, New York, NY 10010, USA\\
	$^2$Max-Planck-Institut f\"ur Astrophysik,
	Karl-Schwarzschild Strasse 1, D-85740 Garching, Germany,\\
	$^3$Department of Physics and Astronomy, Rutgers University, 136 Frelinghuysen Road, Piscataway, NJ 08854, USA\\
	$^4$School of Physics and Astronomy, University of Exeter, Stocker Road, Exeter EX4 4QL, UK
}

\begin{document}
	\maketitle

\begin{abstract}
	We study the destruction of interstellar dust via sputtering in supernova (SN) shocks using three-dimensional hydrodynamical simulations.
	With a novel numerical framework, we follow both sputtering and dust dynamics
	governed by direct collisions, plasma drag and betatron acceleration.
	{Grain-grain collisions are not included and the grain-size distribution is assumed to be fixed.} 
	The amount of dust destroyed per SN is quantified for a broad range of ambient densities and fitting formulae are provided.
	Integrated over the grain-size distribution,
	nonthermal (inertial) sputtering dominates over thermal sputtering for typical ambient densities.
	We present the first simulations that explicitly follow dust sputtering within a turbulent multiphase interstellar medium.
	We find that the dust destruction timescales $\tau$ are 0.35 Gyr for silicate dust and 0.44 Gyr for carbon dust	
	in solar neighborhood conditions.
	The SN environment has an important impact on $\tau$.
	SNe that occur in preexisting bubbles destroy less dust as the destruction is limited by the amount of dust in the shocked gas. 
	This makes $\tau$ about 2.5 times longer than the estimate based on results from a single SN explosion.
	We investigate the evolution of the dust-to-gas mass ratio (DGR), and find that a spatial inhomogeneity of $\sim$ 14\% develops for scales below 10 pc.
	It locally correlates positively with gas density but negatively with gas temperature even in the exterior of the bubbles due to incomplete gas mixing.
	This leads to a $\sim$ 30\% lower DGR in the volume filling warm gas compared to that in the dense clouds.
\end{abstract}

\begin{keywords}
ISM: dust - methods: numerical - Galaxy: solar neighbourhood
\vspace{-0.3cm}
\end{keywords}
\

\vspace{-1.5cm}

\section{Introduction}

Interstellar dust is an important component of galaxies
(for a recent review, see \citealp{2018ARA&A..56..673G} and references therein).
It interacts with stellar radiation via the photoelectric effect, providing an important heating mechanism in the interstellar medium (ISM).
Its surface serves as a site for the formation of molecular hydrogen (often the most efficient channel),
and is a catalyst for the formation of many other molecules. 
Observationally, dust absorbs optical through UV radiation and re-emits it in the IR, thereby modifying the
spectral energy distribution (SED) of galaxies substantially.

Despite its well-recognized importance, 
dust has not received enough attention in most galaxy formation simulations, 
where it is either neglected completely or simply assumed to linearly scale with metals. This is known to be an oversimplification, which breaks down most severely in low metallicity galaxies \citep{2014A&A...563A..31R}.
Very recently,
a few large-scale hydrodynamical simulations \citep{2013MNRAS.432.2298B, 2015MNRAS.449.1625B, 2017MNRAS.468.1505M, 2018MNRAS.478.2851M, 2018MNRAS.478.4905A} as well as semi-analytic models \citep{2017MNRAS.471.3152P} have started to incorporate dust evolution models.
However,
these large scale simulations rely heavily on sub-grid models as they lack the required resolution to model the small-scale grain physics.
Consequently,
these results are subject to uncertain free parameters and the degeneracies they entail.
Our goal in this work is to directly follow the grain physics with high resolution hydrodynamical simulations,
focusing on the destruction processes.

Supernova (SN) shocks are the primary source of dust destruction in the ISM.
The SN blastwaves accelerate and heat up the gas, leading to efficient collisions between dust and gas,
returning the grain material to the gas phase as metals. 
This process, known as sputtering, has been studied extensively in the literature 
\citep{1978ApJ...225..887C, 1978ApJ...226..858S, 1979ApJ...231...77D, 1987ApJ...318..674M, 1989IAUS..135..431M, 1994ApJ...433..797J, 1996ApJ...469..740J, 2014A&A...570A..32B}.
However,
most of the previous studies assumed a steady-state shock model,
whose applicability to realistic time-dependent shocks is uncertain.
\citet{2015ApJ...803....7S} (hereafter SDJ15) took an important step forward:
they conducted a one-dimensional (1D) hydrodynamical simulation of a SN explosion, assuming spherical symmetry,
and then fed the history of the gas into the dust destruction code previously used in \citet{1996ApJ...469..740J}.
This allowed them to study dust destruction in a time-dependent supernova remnant (SNR).
However,
when quantifying the dust destruction timescales,
the limitations of their 1D simulations forced them to make several assumptions about the structure of the multiphase ISM, which is intrinsically three dimensional (3D).

In this work,
we develop and implement a novel numerical framework in a 3D hydrodynamical code, 
allowing us to explicitly simulate the small-scale physics of grain destruction in SN shocks
without resorting to sub-grid models.
This is a first step towards a complete \textit{ab initio} dust evolution model that includes both creation and destruction processes. 
We adopt a one-fluid approach for dust and follow the dust dynamics that is controlled by direct collisions, plasma drag and betatron acceleration.
The dust mass and the relative velocity between dust and gas are integrated in time based on a subcycling technique, 
which can be bypassed when appropriate for the sake of computational efficiency.
We first apply our methods to an idealized problem where a single SN occurs in a uniform medium and quantify the amount of dust destroyed per SN.
We then move on to a more manifestly 3D problem: a multiphase turbulent ISM driven by stochastically injected SNe, which resembles solar-neighborhood conditions.
We quantify the dust destruction timescale and the spatial inhomogeneity of the dust-to-gas mass ratio (DGR) in the multiphase ISM.

This paper is organized as follows:
In Section \ref{sec:method},
we present our numerical framework and then validate our implementation by comparing against analytic solutions.
In Section \ref{sec:singleSN} and \ref{sec:SNbox},
we present our results for dust destruction in single SN exploding in a uniform medium, and multiple SN in a multiphase turbulent ISM, respectively.
We discuss our results in Section \ref{sec:discussion} and summarize our work in Section \ref{sec:summary}.
\section{Numerical Method}\label{sec:method}

\subsection{Hydrodynamics and radiative cooling}

For hydrodynamics,
we use the public version of the {\sc Gizmo} code \citep{2015MNRAS.450...53H},
a multi-method solver based on the meshless Godunov method \citep{2011MNRAS.414..129G} 
and built on the TreeSPH code {\sc Gadget-3} \citep{2005MNRAS.364.1105S}.
We adopt the meshless finite-mass (MFM) solver \citep{2015MNRAS.450...53H} which is a Lagrangian method (i.e. no mass fluxes between particles).
For radiative cooling,
we use the public Grackle chemistry and cooling library \citep{2017MNRAS.466.2217S}\footnote{https://grackle.readthedocs.io/}. 
We adopt the assumption of cooling in ionization equilibrium under a far-UV interstellar radiation field (ISRF) background with a heating rate of 
{$5\times 10^{-26} G_0~ {\rm erg\ s^{-1}}$, 
where $G_0$ is a dimensionless parameter and we set $G_0 = 1$, i.e., the ``Habing field'' \citep{1968BAN....19..421H}.
We do not include the magnetic fields which can provide an extra pressure support.
}
The gas has a metallicity of $Z = 0.02$ (i.e. solar metallicity) which is constant both spatially and temporally\footnote{In fact, the gas-phase metallicity will increase as the dust grains get sputtered and this will in turn change the cooling rates. 
This effect is not included in this paper and is an interesting subject we plan to study in future work.}.

\subsection{Dust model}

We assume that a dust grain is a spherical particle with an internal grain density $\rho_{\rm gr} = 3\ {\rm g/ cm^{3}}$ and grain size (radius) $a$.
We follow two different dust species: carbonaceous dust and silicate dust, which have different erosion rates (see Sec. \ref{sec:sputter}).

\subsubsection{One-fluid approach}

We adopt a one-fluid approach where the dust is spatially coupled with the gas.
This is a good approximation for our purpose
as dust is usually charged and therefore gyrates around the magnetic fields in the ISM with a Larmor radius
\begin{eqnarray}
	r_{\rm L} &=& \frac{m_{\rm gr} {\rm v_{rel}}}{Z_{\rm gr} e B} \nonumber\\
	&\approx& 8.5\times 10^{-3} {\rm pc}\ Z_{\rm gr}^{-1} \Big(\frac{a}{0.1\mu m}\Big)^3 
	\Big(\frac{{\rm v_{rel}}}{{\rm km/s}}\Big) \Big(\frac{B}{3\mu G}\Big)^{-1}
\end{eqnarray}
where $m_{\rm gr} = (4\pi/3)a^3\rho_{\rm gr}$ is the grain mass,
$Z_{\rm gr}$ is the grain charge,
${\rm v_{rel}}$ is the relative velocity (magnitude) between dust and gas,
and $B$ is the magnetic field strength.
Therefore,
the dust is spatially coupled with the magnetic fields and therefore the gas on scales much smaller than 
the structure of the SN shocks ($\sim$ 1 pc),
justifying our assumption.
It should be noted that for micron-size grains ($a \gtrsim 1 \mu m$), 
the Larmor radius can be comparable to or even larger than the scale of the shock structure,
and can therefore escape the magnetic fields and decouple from the gas \citep{2004ApJ...614..796S}.

\subsection{Dust sputtering}\label{sec:sputter}
\begin{figure}
	\centering
	\includegraphics[trim = 10mm 25mm 10mm 35mm, clip, width=0.99\linewidth]{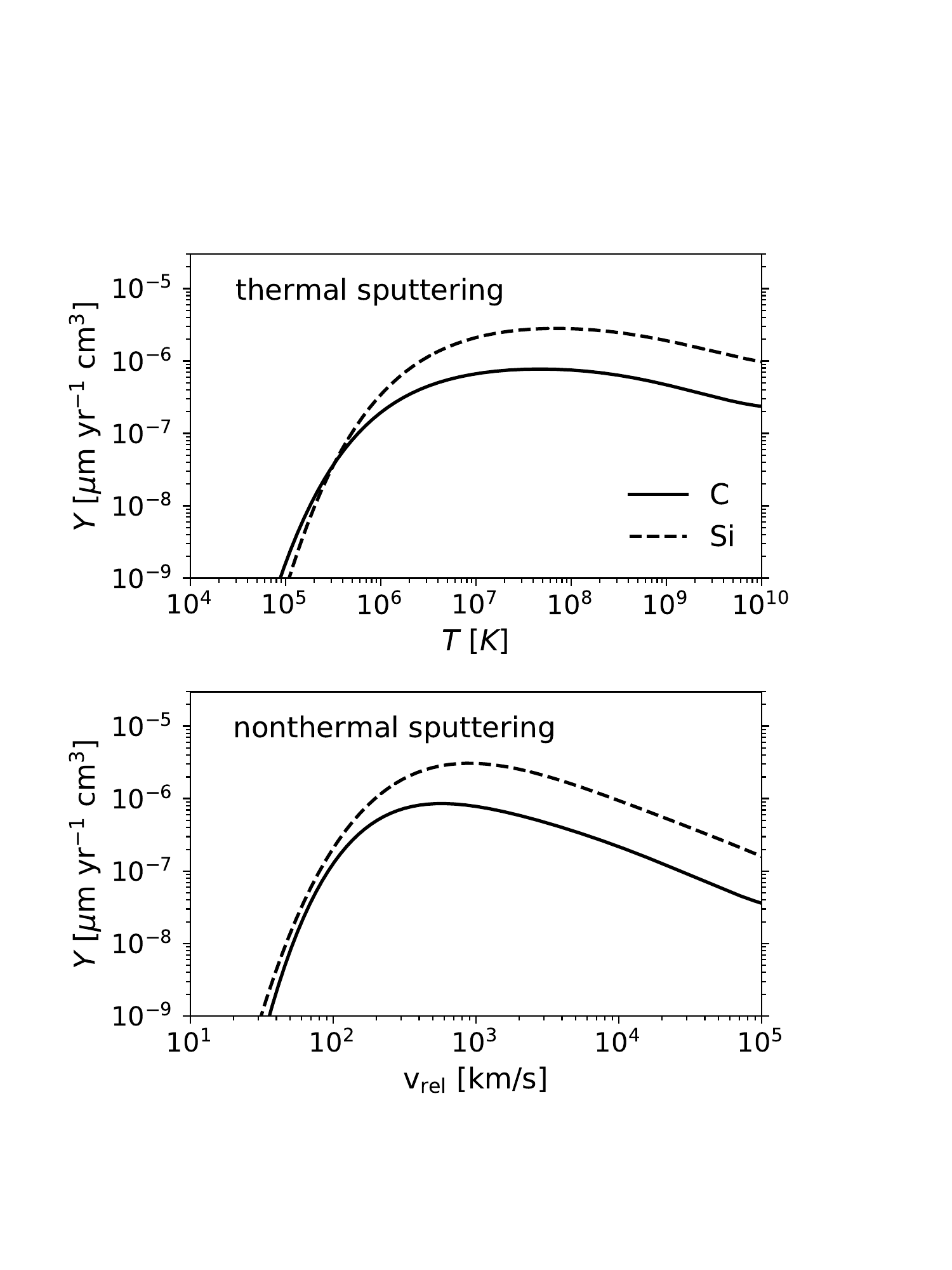}
	\caption{The adopted erosion rates for silicate (dashed line) and carbonaceous (solid line) dust {assuming $10^{-4}$ solar metallicity} from \citet{2006ApJ...648..435N}. \textit{Top}: thermal sputtering as a function of gas temperature ($T$); \textit{Bottom}: nonthermal sputtering as a function of dust-gas relative velocity (${\rm v_{rel}}$). }
	\label{fig:sputY}
\end{figure}

In our one-fluid approach,
each gas particle/parcel has an associated dust mass denoted as $m_{\rm dust}$.
Given a grain size $a$,
the number of grains is then $N_{\rm gr} = m_{\rm dust} / m_{\rm gr}$.
The sputtering rate can be expressed as
\begin{equation}\label{eq:sputeqn}
	 \frac{{\rm d} m_{\rm dust}}{{\rm d} t} = N_{\rm gr}\frac{{\rm d} m_{\rm gr}}{{\rm d} t} = 3 N_{\rm gr}m_{\rm gr} \frac{\dot{a}}{a} = \frac{ 3 n_{\rm H} m_{\rm dust}}{a} Y_{\rm tot},
\end{equation}
where $n_{\rm H}$ is the hydrogen number density of gas
and $Y_{\rm tot}\equiv \dot{a}/n_{\rm H}$ is the ``erosion rates''.
Note that $N_{\rm gr}$ is unaffected by sputtering.
$Y_{\rm tot}$ is the summation of the thermal erosion rates $Y_{\rm th}$ (a function of gas temperature $T$) and the nonthermal (inertial) erosion rates $Y_{\rm nth}$ (a function of ${\rm v_{rel}}$).
We make a polynomial fit to the erosion rates {assuming $10^{-4}$ solar metallicity} from \citet{2006ApJ...648..435N} (their Fig. 2),
adopting their yields of ${\rm MgSiO_4}$ for the silicate dust and C for the carbon dust.
The fitted polynomials can be expressed as $y = \sum_{i=0}^{5} a_i x^i$,
where $x \equiv \log_{10} (T/{\rm K})$ for thermal sputtering (and $x \equiv \log_{10} ({\rm v_{rel}}/ {\rm km\ s^{-1}})$ for nonthermal sputtering),
$y \equiv \log_{10} (Y / {\rm \mu m \ yr^{-1}cm^3})$,
and the $a_i$'s are the coefficients as given in Table \ref{tbl:yields}.
We show the fitted erosion rates for both dust species in Fig. \ref{fig:sputY}.
The sputtering timescale can be defined as 
\begin{eqnarray}
	t_{\rm sput} 
	&\equiv& \frac{a}{3 n_{\rm H} Y_{\rm tot}} \nonumber\\
	&\approx& 0.33 {\rm Myr}\ \Big(\frac{a}{\mu m}\Big) \Big(\frac{n_{\rm H}}{{\rm cm}^{-3}}\Big)^{-1} 
	\Big(\frac{Y_{\rm tot}}{\rm 10^{-6}\mu m \ yr^{-1}cm^3}\Big)^{-1},\nonumber\\ \label{eq:t_sput}
\end{eqnarray}
which can be quite short compared to the dynamical time of the SN shocks, especially for small grains.

Unlike thermal sputtering, which only depends on the local gas properties (i.e. $n_{\rm H}$ and $T$),
nonthermal sputtering also depends on the kinematics of the dust (${\rm v_{rel}}$),
which requires a model of dust dynamics.

{
It is worth noting that the distinction between thermal and nonthermal sputtering is somewhat artificial.
In reality,
the relative velocity between dust and gas is a combination of the drift velocity of dust grains and the thermal motion of gas,
which can be better described as a skewed Maxwellian distribution \citep{1978ApJ...226..858S}.
This more physical distribution has been adopted by \citet{2014A&A...570A..32B} in their steady state calculations.
}

\begin{table}
	\centering
	\begin{tabular}{l | l | lllll}
		\hline	
		\hline	
		& $a_0$ & $a_1$ & $a_2$ & $a_3$ & $a_4$ & $a_5$ \\
		\hline	
		$y_{\rm C, th}$        & -234 & 138   & -33.9 & 4.18 & -0.258 & 0.00639 \\
		$y_{\rm Si, th}$         & -235  & 133  & -31.3 & 3.71 & -0.222 & 0.00532 \\
		$y_{\rm C, nth}$ & -48.7 & 55.7 & -28.8 & 7.44 & -0.965 & 0.005     \\
		$y_{\rm Si, nth}$  & -32.4 & 28.6 & -11.6 & 2.23 & -0.208 & 0.00736 \\
		\hline	
	\end{tabular}
\caption{Polynomial coefficients for our adopted erosion rates in Fig. \ref{fig:sputY},
	where $y = \sum_{i=0}^{5} a_i x^i$,
	$x \equiv \log_{10} (T/{\rm K})$ for thermal sputtering 
	(and $x \equiv \log_{10} ({\rm v_{rel}}/ {\rm km\ s^{-1}})$ for nonthermal sputtering),
	$y \equiv \log_{10} (Y / {\rm \mu m \ yr^{-1}cm^3})$.}
	\label{tbl:yields}
\end{table}

\subsection{Dust dynamics}

Despite the tight spatial coupling between dust and gas due to the gyro-motions,
the relative velocity between dust and gas can be quite high (especially in shocks),
which will in turn lead to nonthermal sputtering.
It is therefore essential to integrate the equation of motion for dust.
Instead of following the ``lab-frame'' dust velocity ${\bf v_{\rm dust}}$,
in our one-fluid approach,
it is more convenient to follow ${\bf v_{\rm rel}} =  {\bf v_{\rm dust}} - {\bf v_{\rm gas}}$ directly,
as the forces acting on both dust and gas (e.g. gravity) will naturally cancel out.
More specifically, 
let ${\bf a_{\rm dust}}$ and ${\bf a_{\rm gas}}$ be the acceleration of dust and gas, respectively,
which can be expressed as
\begin{eqnarray}
	{\bf a_{\rm dust}} &=& {\bf a_{\rm drag}} + {\bf a_{\rm beta}} + {\bf a_{\rm grav}},\\
	{\bf a_{\rm gas}} &=& {\bf a_{\rm hydro}} + {\bf a_{\rm grav}},
\end{eqnarray}
where ${\bf a_{\rm drag}} $ is the acceleration due to the drag force\footnote{We assume that the counteracting acceleration for gas is negligible (i.e. the ``test-particle'' limit) which is appropriate for the typical DGR in the ISM.},
${\bf a_{\rm beta}} $ is the ``betatron acceleration'', which will be described in more detail in Sec. \ref{sec:beta},
and ${\bf a_{\rm grav}}$ and ${\bf a_{\rm hydro}}$ are the acceleration caused by gravity and hydrodynamics, respectively.
Note that ${\bf a_{\rm grav}}$ can be generalized to any acceleration exerted on both dust and gas,
while ${\bf a_{\rm hydro}}$ can be generalized to any acceleration exerted only on gas.
The equation of motion can therefore be expressed as
\begin{eqnarray}\label{eq:xxx}
\frac{\rm d {\bf v}_{\rm rel}}{{\rm d}t} = {\bf a_{\rm drag}} + {\bf a_{\rm beta}} - {\bf a_{\rm hydro}}.
\end{eqnarray}
{For a dust-gas mixture initially at rest which then suddenly encounters a strong shock,
the gas will be accelerated to 0.75 times the shock velocity in the rest-frame (which is naturally captured by the ``$-{\bf a_{\rm hydro}}$'' term in Eq. \ref{eq:xxx}) while the dust is still at rest until the drag force kicks in which gradually drags the dust forward.
}


\subsubsection{Drag force}

The acceleration caused by the drag force can be expressed as \citep{1962pfig.book.....S, 1979ApJ...231...77D, 1987ApJ...318..674M, 2011piim.book.....D}
\begin{eqnarray}\label{eq:drag}
\begin{aligned}
{\bf a}_{\rm drag} &= \frac{{\bf F}_{\rm drag}}{m_{\rm gr}} 
= -\frac{2\pi a^2 n_{\rm H} k_{\rm B} T}{\frac{4\pi}{3}a^3 \rho_{\rm gr}} G({\rm s}) {\bf s} = -\frac{{\bf v_{rel}} }{t_{\rm drag}},
\end{aligned}	
\end{eqnarray}
where $k_{\rm B}$ is the Boltzmann constant, ${\bf s} \equiv {\bf v_{rel}} / (\sqrt{2} c_s)$ and $c_s$ is the speed of sound.
The timescale of the drag force, $t_{\rm drag}$, is defined as
\begin{flalign}
	t_{\rm drag} 
&\equiv& \frac{2\sqrt{2} a \rho_{\rm gr} c_s}{3 n_{\rm H}k_{\rm B} T G({\rm s})} &\nonumber\\
&\approx& 0.59\ {\rm Myr}\ \Big(\frac{a}{\mu m}&\Big) \Big(\frac{n_{\rm H}}{{\rm cm}^{-3}}\Big)^{-1} \Big(\frac{T}{10^6 {\rm K}}\Big)^{-1/2} G(s)^{-1} \label{eq:t_rel},
\end{flalign}
and $G({\rm s})$ is a dimensionless quantity that can be approximated as:
\begin{eqnarray}\label{eq:Gs}
	G({\rm s}) 
	&\approx& \frac{8}{3\sqrt{\pi}}\Big( 1 + \frac{9\pi}{64} s^2 \Big) 
	+ \frac{z^2 \phi^2 \ln(\Lambda/z)}{3\sqrt{\pi}/4 + s^3}\nonumber\\
	&\approx& 1.5 (1 + 0.44 {\rm s}^2)^{0.5} + z^2 \phi^2 \ln(\Lambda/z) (1.3 + {\rm s}^3)^{-1}
\end{eqnarray}
where $z$ is the charge of ions, $\phi$ is the potential parameter, and $\ln(\Lambda/z)$ is the Coulomb logarithm.
The first term in the bracket is due to the \textit{direct collisions} of grains with atoms and ions, while the second term is the \textit{plasma drag} (or dynamical friction) due to Coulomb interaction between the (charged) dust and ions.
The Coulomb potential parameter is defined as
$\phi = Z_{\rm gr} e^2 / (a k_{\rm B}T)$
where $e$ is the electron charge and $Z_{\rm gr}$ is the grain charge.
We adopt a simple recipe for $Z_{\rm gr}$ assuming the dust is photoelectrically charged following the treatment in \citet{2005pcim.book.....T}: 
\begin{eqnarray}
	Z_{\rm gr} \approx 36 \Big(\frac{\gamma}{10^3 {\rm cm^{3} K^{0.5}}}\Big) \Big(\frac{a}{0.1\mu m}\Big),
\end{eqnarray}
where the ionization parameter
$\gamma = G_0 T^{0.5} / n_e$ and $n_e$ is the number density of free electrons.


In a neutral gas,
when the flow is highly subsonic (i.e., $G({\rm s}) \approx$ 1), Eq. \ref{eq:drag} reduces to the \textit{linear} drag relation: ${\bf a}_{\rm drag} \propto {\bf v_{rel}}$,
while when the flow is highly supersonic (i.e., $G({\rm s}) \propto$ s), Eq. \ref{eq:drag} reduces to the \textit{quadratic} drag relation: ${\bf a}_{\rm drag} \propto {\bf v_{rel}}^2$.
However, neither of these are good approximations for the SN shocks
when the gas is ionized with s $\gtrsim 1$.
Therefore, it is necessary to adopt the general form of Eq. \ref{eq:drag}.

\subsubsection{Betatron acceleration}\label{sec:beta}
The charged grains are subject to betatron acceleration \citep{1976ComAp...6..177S, 1978ApJ...226..858S, 1978ApJ...225..887C}, 
which can be expressed as:
\begin{eqnarray}\label{eq:beta}
	{\bf a}_{\rm beta} = \frac{{\bf v_{rel}}}{2 B} \frac{{\rm d} B}{{\rm d} t} \approx \frac{{\bf v_{rel}}}{2\rho_{\rm gas}} \frac{{\rm d}\rho_{\rm gas}}{{\rm d} t} = - \frac{\nabla \cdot {\bf v}_{\rm gas}}{2}{\bf v_{rel}},
\end{eqnarray}
where $B$, $\rho_{\rm gas}$ and $\nabla \cdot {\bf v}$ are the magnetic field strength, density and velocity divergence of gas, respectively.
Following \citet{1979ApJ...231...77D, 1987ApJ...318..674M},
we assume that (i) the magnetic fields are parallel to the shock front and (ii) $B\propto \rho_{\rm gas}$, which are reasonable approximations for strong planar shocks due to flux-freezing\footnote{{As the compression of gas only amplifies the magnetic fields parallel to the shock fronts, the resulting field lines should also be preferentially parallel to the shock fronts. However, this approximation breaks down for the initial magnetic fields normal to the shock fronts.}}.
Note that the normalization of $B$ drops out in Eq. \ref{eq:beta},
as the acceleration comes from the conservation of the magnetic moment $\mu = m_{\rm gr} {\rm v^2_{rel}}/ (2B)$.
{As we do not include magnetic fields in this work,
the betatron acceleration can be overestimated during the shell compression.}

The final equation of motion for dust can be expressed as:
\begin{eqnarray}\label{eq:eom}
	\frac{\rm d {\bf v}_{\rm rel}}{{\rm d}t} = 	- \frac{{\bf v}_{\rm rel}}{t_{\rm rel}} - {\bf a_{\rm hydro}},
\end{eqnarray}
where $t_{\rm rel} \equiv (t_{\rm drag}^{-1}  - (\nabla \cdot {\bf v}_{\rm gas})/2)^{-1}$.

\subsection{Time-integration scheme}

Eq. \ref{eq:sputeqn} and \ref{eq:eom} have to be coupled during the integration because nonthermal sputtering depends on ${\rm v_{rel}}$.
In addition,
the two equations also have to be coupled with the radiative cooling,
as both of them depend on temperature.
However,
the timescales $t_{\rm sput}$ and $t_{\rm rel}$ can be much smaller than the typical hydrodynamical timesteps $\Delta t$ in SN shocks.
This is especially true for the small grains which couple tightly with the gas.
In this case,
explicit time integration schemes would lead to numerically unstable solutions.

Implicit methods have been commonly adopted to ensure numerical stability while taking a large timestep (e.g. \citealp{2010ApJS..190..297B}).
Alternatively, semi-analytic methods \citep{2015MNRAS.452.3932B, 2015MNRAS.454.4114L} can also be used to ensure numerical stability,
though these can only be used where analytic solutions are available (e.g. for the linear drag relation),
which is not applicable in SN shocks (see Eq. \ref{eq:drag}).
Moreover,
despite being numerically stable,
these methods are not suitable for our purpose,
which requires \textit{accurate} integration throughout a hydrodynamical timestep (rather than just the terminal values at the end of a timestep).

We therefore adopt an integration scheme based on sub-cycling, a
technique frequently adopted to solve radiative cooling and chemistry
reactions, to tackle the stiffness problem.  A hydrodynamical timestep
is divided into multiple sub-timesteps.  Within each sub-timestep, we
integrate the equation using an explicit 2nd-order predictor-corrector
method (i.e. a midpoint rule).  Namely, for an equation $\dot{y} = f
(y)$, we update $y$ as follows:
\begin{eqnarray}
	\tilde{y}^{n+1}     &=& y^n + \dot{y}^n\  \Delta t_{\rm sub}, \nonumber\\ 
	\dot{y}^{n+1} &=& f (\tilde{y}^{n+1}  ), \\
	y^{n+1}     &=& y^n + 0.5\ (\dot{y}^n + \dot{y}^{n+1})  \Delta t_{\rm sub}, \nonumber
\end{eqnarray}
where $n$ is the discretized time integer and $\Delta t_{\rm sub}$ is the sub-timestep determined by $t_{\rm sub} = f_{\rm sub} \min({t_{\rm rel}}, {t_{\rm sput}}, {t_{\rm c}})$, with $t_{\rm c}$ the local cooling time. 
The quantity $f_{\rm sub}$ is a sub-cycling factor which controls the desired accuracy.
Note that $t_{\rm sub} $ is updated at the beginning of every sub-timestep.

\begin{figure}
	\centering
	\includegraphics[trim = 0mm 5mm 0mm 0mm, clip, width=0.9\linewidth]{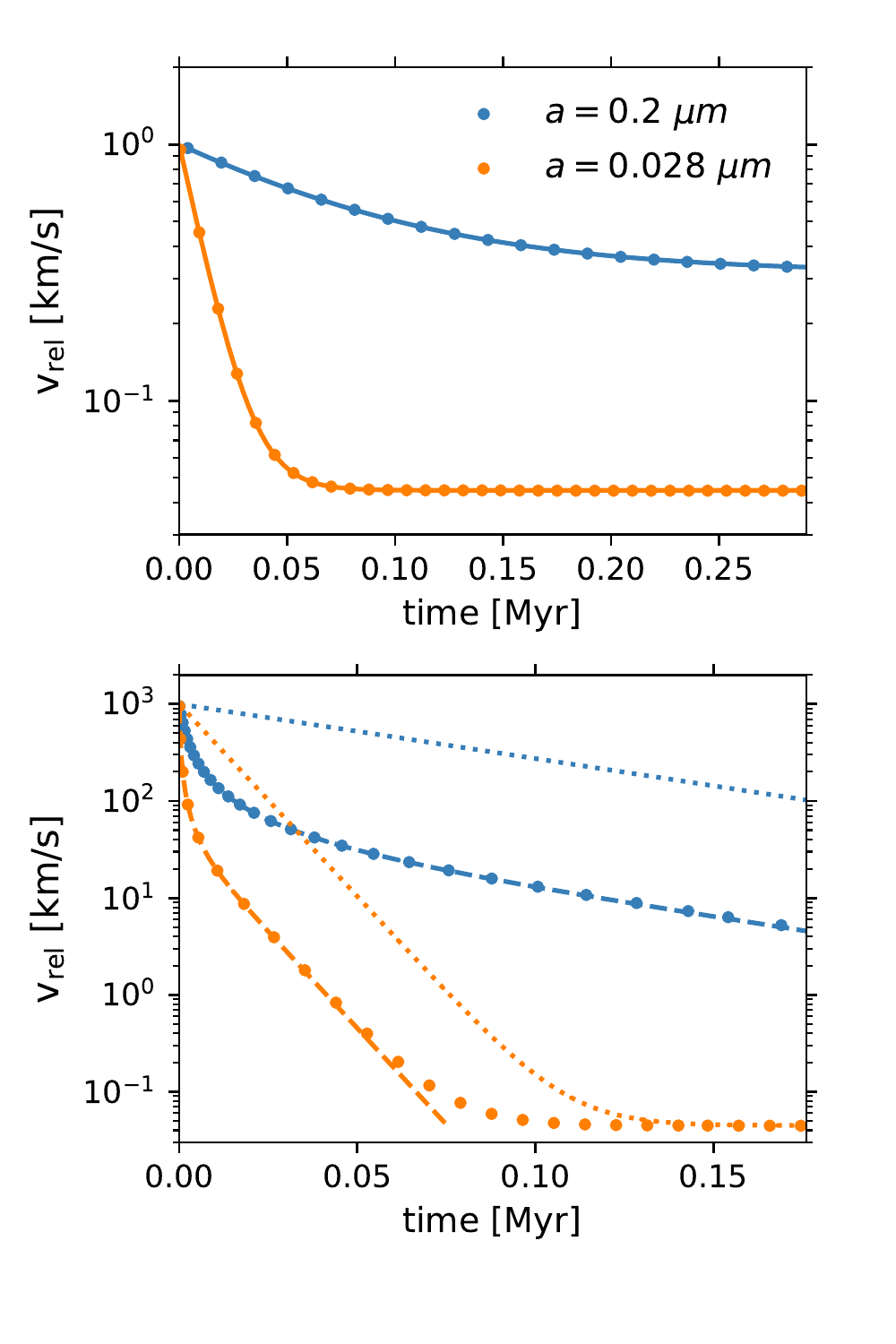}
	\caption{\textit{Upper panel:} time evolution of ${\rm v_{rel}}$ with an initial value of 1 km/s in a neutral medium of $n_{\rm H} = 10 {\rm cm^{-3}}$ and $T = 10^4$K with two different grain sizes $a = 0.2\mu m$ (in blue) and $a = 0.028\mu m$ (in orange).
		The gas is initially at rest ($v_{\rm gas} = 0$) but is subject to a constant acceleration $a_{\rm hydro} = -4000\, {\rm kms^{-1} Gyr^{-1}}$.
		The circles are the numerical results of our time integration scheme which agree very well with the analytic solutions (solid lines).
		\textit{Lower panel:} same setup but with ${\rm v_{rel}} = 10^3 \, {\rm km/s}$ initially, which is in the supersonic regime.
		Analytic solutions for $a_{\rm hydro}=0$ are shown as the dashed lines, while solutions that (incorrectly) assume a linear drag relation are shown as the dotted lines.
	}
	\label{fig:dragtest}
\end{figure}

\subsubsection{Validation}
\begin{figure}
	\centering
	\includegraphics[trim = 5mm 0mm 5mm 5mm, clip, width=0.99\linewidth]{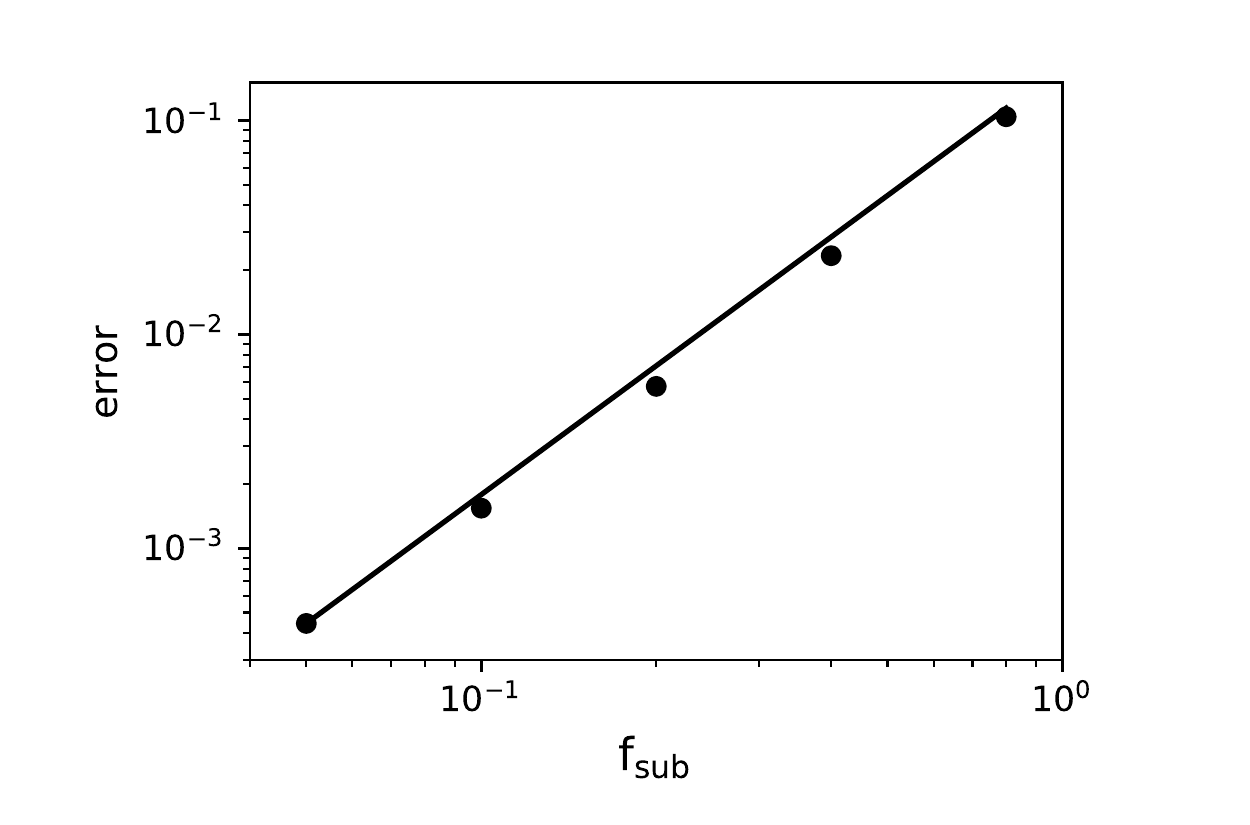}
	\caption{Numerical error (circles) as a function of $f_{\rm sub}$.
		The error, defined as the time-averaged $|{\rm v_{rel}} - {\rm v_{rel,true}}| / {\rm v_{rel,true}}$ where ${\rm v_{rel,true}}$ is the analytic solution,
		scales as $f_{\rm sub}^2$ (solid line),
		implying that our method is indeed second order accurate.
	}
	\label{fig:dragtestconv}
\end{figure}

In the upper panel of Fig. \ref{fig:dragtest},
we show the time evolution of ${\rm v_{rel}}$ with an initial value of 1 km/s in a neutral medium with density $n_{\rm H} = 10 {\rm cm^{-3}}$ and $T = 10^4$K, with two different grain sizes $a = 0.2\mu m$ (in blue) and $a = 0.028\mu m$ (in orange).
The gas is initially at rest ($v_{\rm gas} = 0$) but is subject to a constant acceleration $a_{\rm hydro} = -4000\, {\rm km s^{-1} Gyr^{-1}}$.
The circles are the numerical results of our time integration scheme which agree very well with the analytic solutions (solid lines).
The drag force in this case is in the subsonic regime ($s\ll 1$) and a linear drag relation applies.
The solution is a simple exponential decay until settling to a terminal velocity $a_{\rm hydro} t_{\rm vel}$ (see Eq. \ref{eq:linearV}).
In the lower panel of Fig. \ref{fig:dragtest}, we show the time evolution of the same setup but with ${\rm v_{rel}} = 10^3\, {\rm km/s}$ initially,
which is in the supersonic regime ($s\gg 1$).
Analytic solutions for this setup are available only for $a_{\rm hydro}=0$,
which is shown as the dashed lines (cf. Eq. \ref{eq:nonlinearV}).
Our numerical results agree well with the analytic solutions until settling to the terminal velocity,
which is not captured by the analytic solution.
We emphasize that it is important to adopt a scheme that can correctly follow the supersonic drag force.
For example,
semi-analytic methods such as those proposed by \citet{2015MNRAS.452.3932B} and \citet{2015MNRAS.454.4114L}, which only apply for the subsonic cases, will predict solutions that significantly underestimate the drag force, as shown in Fig. \ref{fig:dragtest}.
In Fig. \ref{fig:dragtestconv},
we show the numerical error as a function of $f_{\rm sub}$.
The error, defined as the time-averaged $|{\rm v_{rel}} - {\rm v_{rel,true}}| / {\rm v_{rel,true}}$ where ${\rm v_{rel,true}}$ is the analytic solution,
scales as $f_{\rm sub}^2$,
implying that our method is indeed second order accurate.

\subsubsection{Sub-cycling switch}
Although sub-cycling solves the stiffness problem,
it can still be computationally expensive when the number of sub-cycles is large.
More importantly,
there are situations where sub-cycling is unnecessary for our purpose (e.g. integrating ${\rm v_{rel}}$ when it is essentially zero).
Thermal sputtering is negligible when $T < 10^5$K while nonthermal sputtering is negligible when ${\rm v_{rel}} < 30 {\rm km/s}$ (cf. Fig. \ref{fig:sputY}). 
Furthermore,
if ${\rm v_{rel}} < 30 {\rm km/s}$,
the integration of ${\rm v_{rel}}$ and $m_{\rm dust}$ can be decoupled
(though they still need to be coupled with radiative cooling due to their $T$-dependence).
In this case,
we only need to ensure that the final ${\rm v_{rel}}$ is integrated correctly at the end of the timestep.
This leads to alternative integration schemes that are much more efficient than sub-cycling.
In the sub-sonic regime ($s < 0.5$) where the drag relation is linear\footnote{For $s=0.5$, the nonlinear terms for direct collisions and the plasma drag (cf. Eq. \ref{eq:Gs}) are about 5\% and 10\%, respectively, sufficiently small to be neglected.},
we can integrate ${\rm v_{rel}}$ with the exact solution following \citet{2015MNRAS.452.3932B, 2015MNRAS.454.4114L}, i.e.
 \begin{eqnarray} \label{eq:semianaly}
 	{\rm v^{n+1}_{rel}} = {\rm v^n_{rel}} \exp(-t / t_{\rm rel}) - a_{\rm hydro} t_{\rm rel} (1 - \exp(-t / t_{\rm rel})  ),
 \end{eqnarray}
which is capable of capturing the terminal velocity ${\rm v^{term}_{rel}} = -a_{\rm hydro} t_{\rm rel}$.
Moreover, 
if $\Delta t \gg t_{\rm rel} $,
we can update ${\rm v_{rel}}$ directly with ${\rm v^{term}_{rel}}$,
and we do so when $\Delta t > 3t_{\rm rel}$ as the velocity has decayed by a factor of $\e^{-3} \approx 5\%$.

Our algorithm can be summarized as follows:
\begin{algorithmic}
	\If {${\rm v_{rel}} < 30\, {\rm km/s}$}
	\State \textit{/* $T ~\&~  {\rm v_{rel}}$ can be decoupled */}
		\If {$T > 10^5$K}
			\State $sub[ (m_{\rm dust}, T), \min(t_{\rm sput}, t_{\rm cool}) ]$
		\EndIf
		\If {$ \Delta t > 3\ t_{\rm rel} $} 
		\State \textit{/* approach terminal velocity */}
			\State $ {\rm v_{rel}}  \gets {\rm v^{term}_{rel}}$
		\Else {}
			\If {$s < 0.5$} 
			\State \textit{/* analytic solution available */}
				\State $update\ {\rm v_{rel}}\ with$ Eq. \ref{eq:semianaly}
			\Else
				\State $sub[ (m_{\rm dust}, T), \min(t_{\rm rel}, t_{\rm cool}) ]$
			\EndIf
		\EndIf
	\Else 
		\State \textit{/* fully coupled integration */}
		\State $sub[ (m_{\rm dust}, {\rm v_{rel}}, T), \min(t_{\rm sput}, t_{\rm rel}, t_{\rm cool}) ]$
	\EndIf
\end{algorithmic}
where $sub[(y_1, y_2), \delta t]$ refers to integrating the variables $y_1$ and $y_2$ with a sub-timestep of $\delta t$.
Note that we always need to integrate $T$ during the sub-cycling as the sputtering and dust dynamics both depend on $T$.
When $m_{\rm dust}$ and ${\rm v_{rel}}$ are sub-cycled separately,
$T$ needs to be reset after the first round of sub-cycling to avoid double-counting the cooling.

\subsection{Grain-size distribution}

Sputtering and dust dynamics both depend sensitively on the grain size (cf. Eq. \ref{eq:t_sput} and \ref{eq:t_rel}).
We assume that the probability distribution function of grain size is $f(a)\propto a^{-3.5}$ in the range of $[a_{\rm min}, a_{\rm max}] = [0.005, 0.25] \mu m$,
which is the so-called the MRN distribution \citep{1977ApJ...217..425M} applicable for Milky Way-like dust in the diffuse ISM.

\subsubsection{Constant distribution approximation}\label{sec:size}

As dust gets sputtered, 
the size distribution will evolve.
Sputtering reduces the grain size and leads to mass flux towards smaller grain size bins.
The distribution also evolves due to the grain size dependent sputtering rate.
In addition,
grain shattering, which is not included in our model, would create fragmentation of grains.
The net effect of SNe is therefore to transform large grains into small grains, changing the size distribution,
which becomes a runaway process if we only consider SNe.
Following the evolution of the grain size distribution self-consistently in a SNR is possible and in fact has been done in previous studies (e.g. \citealp{1996ApJ...469..740J}, \citealp{2014A&A...570A..32B} and SDJ15).
However, 
the assumption is that every SN will process dust that has the same initial size distribution,
which is not entirely self-consistent as the same dust can be processed by several SNe.
In fact, 
the implicit assumption is that dust processed by one SN will have its size distribution readjusted back to its initial distribution before the next SN arrives.
{The physical justification is that the size distribution is in a steady state in a statistical sense,
which can only be realized if physical processes besides SNe that can replenish large grains (e.g. in the stellar ejecta of asymptotic giant branch stars) and transform small grains into large grains (e.g. ISM growth and coagulation) are present.
Such fully self-consistent modeling is beyond the scope of this paper and we leave it to future work.
On the other hand,
in hydrodynamical simulations of multiple SNe (as will be shown in Sec. \ref{sec:SNbox}),
SNRs overlap and interact with each other in a nontrivial way and readjusting the size distribution after the processing of each SN is no longer a well-defined procedure.
}
In this work,
we simply make the assumption that the size distribution always remains constant due to other dust processes that we do not model explicitly. 
\citet{1996ApJ...469..740J} have shown that shattering naturally generates fragments which roughly follow the power law distribution close to MRN,
providing some physical justification for our approach.
{In addition,
\citet{2014A&A...570A..32B} found that ignoring grain-grain collisions only slightly decreases the final sputtered mass ($\sim$3\%),
even if the size distribution is broadened by shattering.
}

\begin{figure*}
	\centering
	\includegraphics[trim = 10mm 5mm 10mm 0mm, clip, width=0.99\linewidth]{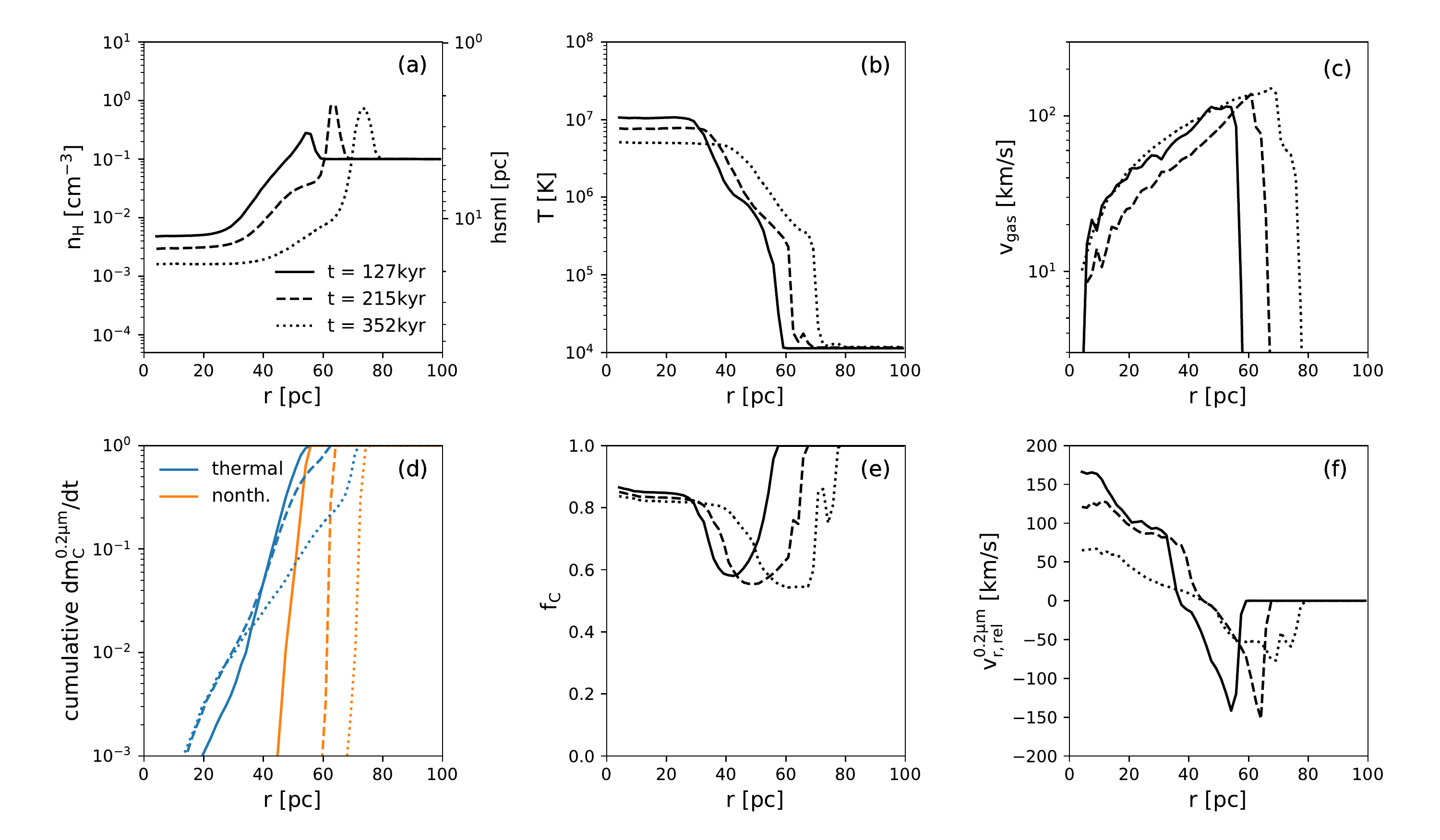}
	\caption{
		Radial profile of the supernova remnant (SNR) in an initially uniform medium of $n_{\rm H,0} = 0.1 {\rm cm^{-3}}$ at time $t$ = 127 (solid), 215 (dashed) and 352 kyr (dotted), respectively. The cooling time for the SNR is $t_c$ = 156 kyr.
		{Panel (a)}: gas hydrogen number density $n_{\rm H}$ {(left axis) and the corresponding smoothing length $h$ (right axis)};
		{panel (b)}: gas temperature $T$;	
		{panel (c)}: gas radial velocity $v_{\rm gas,r}$;
		{panel (d)}: cumulative normalized distribution of the rate of thermal (orange) and nonthermal (blue) sputtering for the carbon dust for $a = 0.2\mu m$ (see text for definition);
		{panel (e)}: mass fraction of the carbon dust relative to its initial value $f_{\rm C}$;
		{panel (f)}: dust-gas relative velocity in the radial direction ${\rm v_{r,rel}}$ for $a = 0.2\mu m$.
	}
	\label{fig:fig1profile}
\end{figure*}

\subsubsection{Discrete formulation}
Given the dust mass associated with a gas particle $m_{\rm dust}$,
we discretize it into $N_{\rm bin}$ logarithmic bins.
The $a$-bins are defined as:
\begin{eqnarray}
da &=& (\log_{10}a_{\rm max} - \log_{10}a_{\rm min}) / N_{\rm bin}, \\
\log_{10}{a_i} &=& \log_{10}{a_{\rm min}} + (i-0.5)da, \\
\log_{10}{a_{i\pm1/2}} &=& \log_{10}{a_{\rm min}} + (i-0.5\pm0.5)da,
\end{eqnarray}
where $a_i$ is the central value in the $i$-th bin while $a_{i\pm1/2}$ are the boundaries of the $i$-th bin.
We choose $N_{\rm bin}=8$ throughout this paper.
Defining the $n$-th moment of $f(a)$ as
\begin{equation}
	\langle a^n\rangle_i  \equiv \int^{a_{i+1/2}}_{a_{i-1/2}} a^n f(a) da ,
\end{equation}
the mass fraction in the $i$-th bin can be pre-calculated analytically as
${\langle a^3\rangle_i} / { \sum_i  \langle a^3\rangle_i}$.
The dust mass in the $i$-th bin can be expressed as
$m^{i}_{\rm dust} = \langle a^3\rangle_i$\footnote{Technically, it should be $\langle (4\pi/3)\rho_{\rm gr}a^3\rangle_i$. However, the prefactor $(4\pi/3)\rho_{\rm gr}$ can be absorbed into the normalization of $f(a)$, simplifying the expression.}
and its time derivative becomes
\begin{equation}
	\dot{m}^{i}_{\rm dust} = \langle 3a^2 \dot{a}\rangle_i \approx 3\dot{a}_i  \langle a^2\rangle_i \,
\end{equation}
where we made an approximation that $\dot{a}$ is independent of $a$ \textit{within} the $i$-th bin\footnote{
For thermal sputtering,
this is in fact not an approximation but holds strictly true.
For nonthermal sputtering,
$\dot{a}$ depends on $a$ as the drag force is a function of $a$,
so we are effectively assuming that ${\rm v_{rel}}$ is piecewise constant within the $i$-th bin.
}.
We can then write down the sputtering equation in a similar form as the single grain-size version (Eq. \ref{eq:sputeqn}):
\begin{equation}
	\dot{m}^{i}_{\rm dust} = 3 \dot{a}_i \frac{\langle a^2\rangle_i}{\langle a^3\rangle_i} m^{i}_{\rm dust}
	 = 3 \frac{\dot{a}_i }{a_i^{\rm eff}} m^{i}_{\rm dust},
\end{equation}
where $a_i^{\rm eff}\equiv {\langle a^3\rangle_i} / {\langle a^2\rangle_i} $ is the effective grain-size. 
As we assume a power law distribution for $f(a)$,
the effective grain-size can be computed exactly:
\begin{equation}
	a_i^{\rm eff} = \frac{(a_{i+1/2}^{0.5} - a_{i-1/2}^{0.5})/0.5}{(a_{i+1/2}^{-0.5} - a_{i-1/2}^{-0.5})/(-0.5)}
	 = (a_{i+1/2}a_{i-1/2})^{0.5}, 
\end{equation}
which happens to be the geometric mean of the boundaries and has the nice property that $\log_{10}a_i^{\rm eff}$ coincides with the central value $\log_{10}a_i$ \footnote{This is a coincidence due to the specific choice of the power-law index $f(a)\propto a^{-3.5}$. Choosing a power-law index other than -3.5 will make $a_i^{\rm eff}$ deviate from the geometric mean of the boundaries, though it can still be analytically computed.}.
The time integration scheme described in the previous sections is applied to each bin separately:
{each bin follows its own dust-gas relative velocity and dust mass.
At the end of a timestep,
the total (grain-size integrated) dust mass is decreased according to the total sputtering rate while the grain-size distribution remains the same.}


\subsection{Terminology}
For the sake of clarity,
we define a few pieces of terminology that will be used frequently throughout the paper.
The dust mass associated with each particle is denoted as $m_{\rm C}$ for the carbon dust and $m_{\rm Si}$ for the silicate dust. 
We will describe our definitions and notation for the carbon dust as an example,
while the same terms apply for the silicate dust by replacing the underscript $(\ )_{\rm C}$ with $(\ )_{\rm Si}$.
The DGR of each gas particle is denoted as
\begin{eqnarray}
D_{\rm C} = m_{\rm C} / m_{\rm gas}.
\end{eqnarray}
The ratio of current dust mass to its initial value is defined as
\begin{eqnarray}
f_{\rm C}   \equiv m_{\rm C} / m_{\rm 0,C} = D_{\rm C} / D_{\rm 0,C},
\end{eqnarray}
where the subscript $(\ )_{\rm 0}$ indicates the values in the initial conditions.
Note that $0\leq f_{\rm C}\leq 1$ as we do not include any mechanisms for dust creation in this work.
Similarly,
we define the ratio of the total dust mass of the entire system to its initial value as
\begin{eqnarray}
F_{\rm C}   \equiv \frac{ \sum_i m^i_{\rm C} }{ \sum_i m^i_{\rm 0,C} }= \frac{ \sum_i D^i_{\rm C} }{ \sum_i D^i_{\rm 0,C} },
\end{eqnarray}
where the index $i$ refers to the $i$-th particle in the system and the summation is over all particles.
Since we will always adopt equal-mass gas particles and a spatially uniform DGR in the initial conditions,
we can drop the $i$-index for $m_{\rm 0,C}$ and $D_{\rm 0,C}$,
and the global DGR of the system can be expressed as 
\begin{eqnarray}
\overline{D}_{\rm C} = \frac{\sum_i m^i_{\rm C} }{ N_{\rm gas} m_{\rm gas}} =  \frac{\sum_i D^i_{\rm C} }{N_{\rm gas}}.
\end{eqnarray}
The \textit{sputtered mass} is defined as the total amount of dust mass returned back to the gas phase by sputtering and is expressed as
\begin{eqnarray}
M_{\rm sp,C}   \equiv m_{\rm 0,C} \sum_i (1 - f^i_{\rm C}),
\end{eqnarray}
which explicitly depends on the adopted $D_{\rm 0,C}$.
To factor out the dependence on $D_{\rm 0,C}$,
we define the \textit{gas mass cleared of dust} as 
\begin{eqnarray}
M_{\rm cl,C} \equiv M_{\rm sp,C} / D_{\rm 0,C} = m_{\rm gas} \sum_i (1 - f^i_{\rm C}).
\end{eqnarray}
Finally,
the \textit{sputtering efficiency} is defined as in \citet{1980ApJ...239..193D}:
\begin{eqnarray}
\epsilon_{\rm sp,C} \equiv M_{\rm cl,C}  / M_{\rm swept},
\end{eqnarray}
where $M_{\rm swept}$ is the gas mass swept up by the SN shock,
measured by the total amount of gas mass that reaches a gas velocity $v_{\rm gas} > 30 {\rm km/s}$.
Unless otherwise stated,
all quantities are grain-size integrated.
We use a superscript $(\ )^{a}$ to refer to properties associated with a given $a$-bin,
e.g., $M^a_{\rm sp,C}$.
When referring to quantities related to thermal or nonthermal sputtering,
we use a superscript $(\ )^{\rm th}$ or $(\ )^{\rm nth}$.

\section{Single SN in uniform medium}\label{sec:singleSN}

In this section,
we investigate dust destruction in a SNR occurring in an initially uniform and static medium.
The initial hydrogen number density of the background medium is $n_{\rm H,0}$, which we will vary systematically.
We choose an initial gas temperature of $T_0 = 10^4$K and initial DGR $D_{\rm C} = D_{\rm Si} = 0.005$.
The particle mass of the gas is $m_{\rm gas} = 0.04 {\rm M_\odot}$
(our convergence study in Appendix \ref{app:convtest} suggests that the results converge at $m_{\rm gas} = 0.2 {\rm M_\odot}$).
We adopt the canonical supernova explosion energy $E_{\rm SN} = 10^{51}$erg \citep{1999ApJS..123....3L},
which is injected into $N_{\rm inj} = \max(32, 16{\rm M_\odot}/m_{\rm gas}) = 400$ nearest gas particles in a kernel-weighted fashion.
This particular choice of $N_{\rm inj}$ is to make sure that the injection mass $N_{\rm inj}m_{\rm gas} = 16 {\rm M_\odot}$ is close to the mass of the SN ejecta,
which marks the transition from the free-expansion phase to the Sedov-Taylor phase.
When $m_{\rm gas}$ is too large to resolve $16 {\rm M_\odot}$,
we choose $N_{\rm inj}=32$ which is about one resolution element.

\subsection{Time evolution}

\begin{figure*}
	\centering
	\includegraphics[trim = 20mm 10mm 20mm 10mm, clip, width=0.99\linewidth]{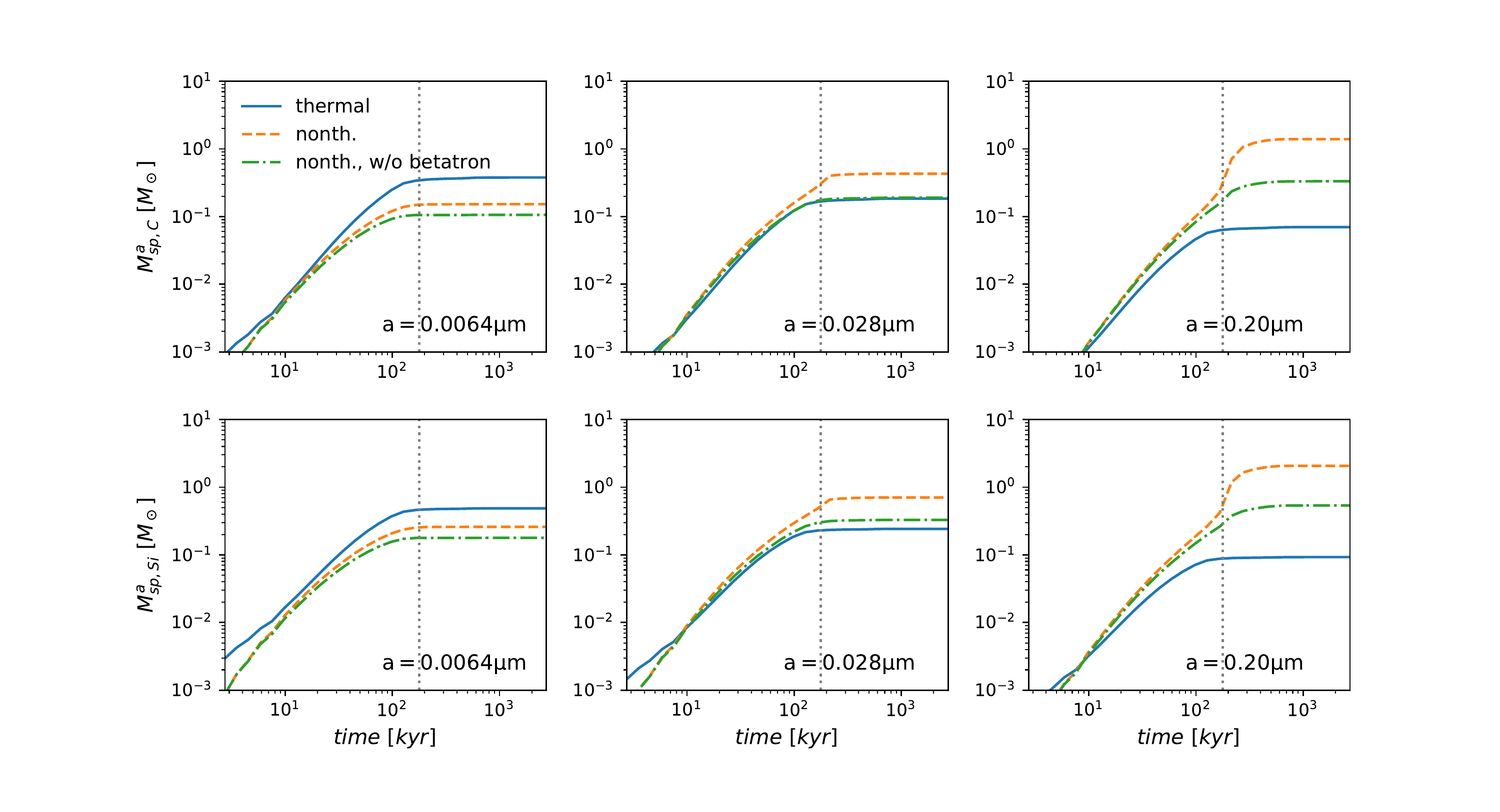}
	\caption{Time evolution of $M^{a}_{\rm sp,C}$ (upper panels) and $M^{a}_{\rm sp,Si}$ (lower panels) for three different grain sizes $a = 0.0064\mu m$ (left), $0.028\mu m$ (middle) and $0.2\mu m$ (right), respectively.
		The blue solid line is for thermal sputtering and the orange dashed line is for nonthermal sputtering.
		The green dash-dotted line is for nonthermal sputtering but without betatron acceleration. 
		The initial gas density is $n_{\rm H,0} = 0.1 {\rm cm^{-3}}$, and the vertical grey dotted line indicates the cooling time $t_c = 156$ kyr.	
		Thermal sputtering dominates for small grains and operates only before $t_c$.
		Nonthermal sputtering dominates for large grains and becomes most efficient right after $t_c$ due to betatron acceleration.
	}
	\label{fig:fig2_Msp_vs_time}
\end{figure*}

\subsubsection{Radial profile}
In Fig. \ref{fig:fig1profile},
we show the radial profile of the SNR for $n_{\rm H,0} = 0.1 {\rm cm^{-3}}$ at three different times $t$ = 127 (solid), 215 (dashed) and 352 kyr (dotted), respectively.
The cooling time for the SNR is $t_c$ = 44 kyr$(n_{\rm H,0}/{\rm cm^{-3}})^{-0.55}$ = 156 kyr \citep{2015ApJ...802...99K}. 
The evolution of gas properties in a SNR have been studied extensively with hydrodynamical simulations in the literature,
both in 1D spherical symmetric codes \citep{1972ApJ...178..159C, 1974ApJ...188..501C, 1988ApJ...334..252C, 1998ApJ...500...95T}
and more recently in several different 3D codes 
(\citealp{2015ApJ...809...69S, 2015ApJ...802...99K, 2015MNRAS.450..504M, 2015MNRAS.451.2757W, 2016MNRAS.460.2962H}, and see \citealp{2017ARA&A..55...59N} for a review).
Our results for gas properties (panels a, b and c) broadly agree with previous works.
Before $t=t_c$,
the SNR is in the Sedov-Taylor phase, where the total energy is conserved and the radial momentum increases with time as the over-pressurized gas drives the expansion of the spherical shock front.
In this phase,
the theoretical maximum compression factor of the shock front is $\chi = 4$, limited by the pressure support of the shell,
and we find $\chi \sim 3$ due to our finite resolution.
After $t_c$,
the SNR gradually loses energy due to radiative cooling in the shell,
causing further compression of the shell up to $\chi \sim 10$ in our case.


As it is a collision process, 
sputtering is most efficient when the gas is both dense and hot.
In a SNR,
the gas temperature in the diffuse bubble is usually higher than that in the dense shell.
Therefore, 
it is not obvious which of the two regions will experience the most dust destruction.
In panel (d),
we show the cumulative distribution of the sputtering rate for the carbon dust for $a = 0.2\mu m$ (normalized to one) defined as
$\sum_{i}^{r_i<r} \dot{m}_{{\rm C},i}^{0.2\mu m}  / \sum_{i} \dot{m}_{{\rm C},i}^{0.2\mu m}$.
Thermal and nonthermal sputtering are denoted as orange and blue lines, respectively.
It can be seen that most of the dust destruction happens in the dense shell rather than the diffuse hot bubble,
because (i) collision is more efficient in denser gas (cf. Eq. \ref{eq:sputeqn}), (ii) most of the mass is concentrated in the dense shell, and (iii) ${\rm v_{rel}}$ increases due to betatron acceleration after $t_c$ which occurs mainly in the shell (relevant for nonthermal sputtering).
The distributions for smaller grains (not shown) follow a similar trend that most dust destruction happens in the dense shell rather than in the diffuse bubble.
{The Lagrangian nature of our simulations is well-suited to resolve the dense shell where most dust sputtering occurs.
While the spatial resolution in the diffuse bubble becomes more coarse, dust sputtering in that region is negligible. }

Panel (e) shows the radial profile of $f_{\rm C}$,
which reaches a minimum around 0.6 (i.e. 40\% of the dust is destroyed).
In the central region,
$f_{\rm C}$ is actually higher as the sputtering rate in the diffuse bubble is low.
The location of minimal $f_{\rm C}$ slightly lags behind the shell because the sputtering timescale is not much shorter than the dynamical timescale of the SNR.
Therefore,
there is a finite time delay between the gas being shocked and the dust being destroyed.

Panel (f) shows ${\rm v_{rel}}$ in the radial direction as a function of radius for $a = 0.2\mu m$,
which reaches a peak value right after the shock front and then is gradually decelerated by the drag force and eventually overshoots in the diffuse bubble.
After $t_c$,
the effect of betatron acceleration can be seen clearly at $t = 215$ kyr where the peak ${\rm v_{rel}}$ exceeds the radial gas velocity (panel c) at the same radius.
Smaller grains (not shown) have lower ${\rm v_{rel}}$ as they couple more tightly to the gas but the general trend is similar.

\subsubsection{Sputtered mass and sputtering efficiency}
In Fig. \ref{fig:fig2_Msp_vs_time},
we show the time evolution of $M^{a}_{\rm sp,C}$ and $M^{a}_{\rm sp,Si}$ for different grain sizes ($a = 0.0064\mu m$, $0.028\mu m$ and $0.2\mu m$).

\begin{figure}
	\centering
	\includegraphics[trim = 0mm 20mm 0mm 10mm, clip, width=0.99\linewidth]{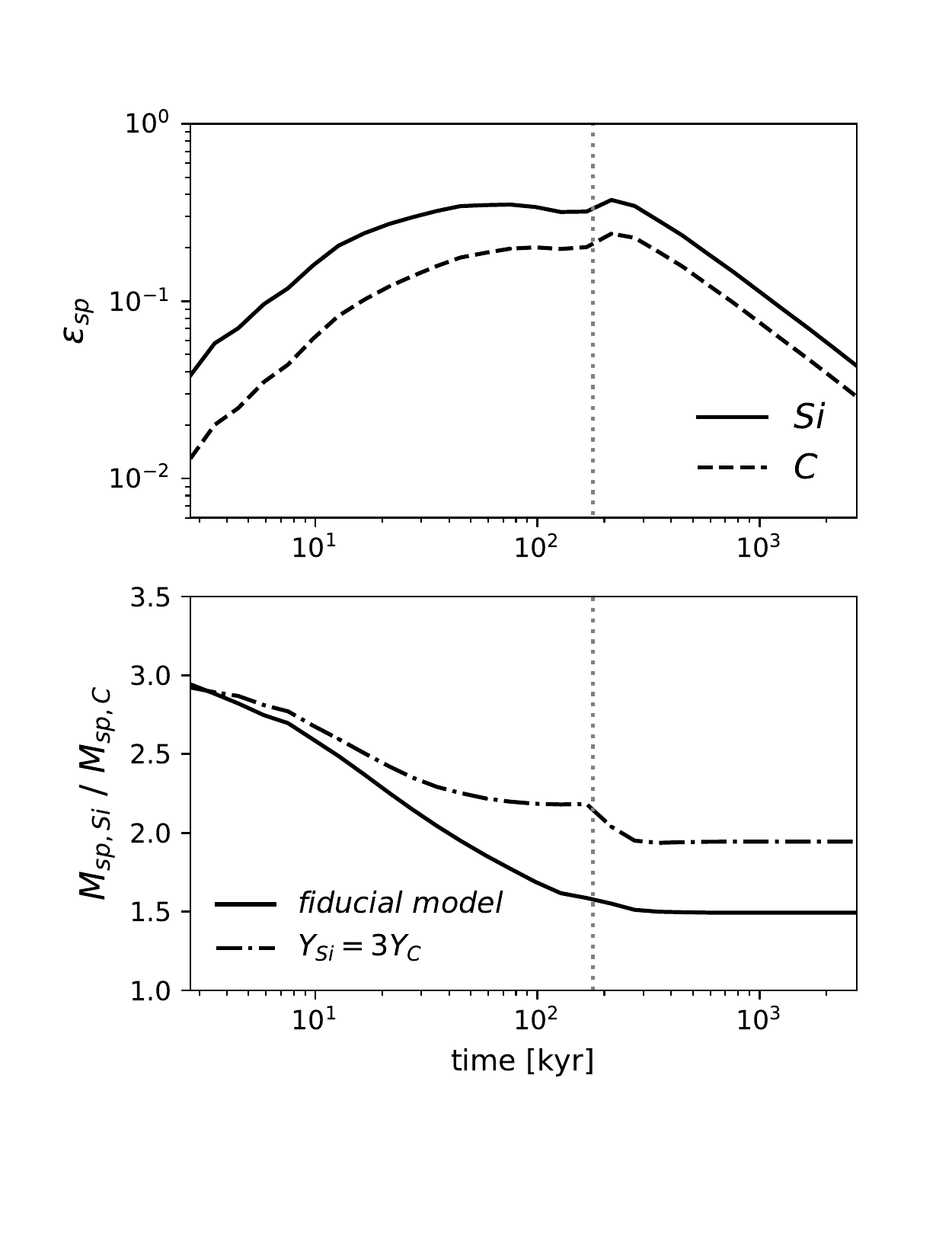}
	\caption{
		\textit{Upper panel}: time evolution of the sputtering efficiency for the carbon dust (dashed line) and silicate dust (solid line), respectively. 
		\textit{Lower panel}: time evolution of the ratio of $M_{\rm sp,Si}$ to $M_{\rm sp,C}$ (solid line).  The dash-dotted line shows the same ratio but in a model where the erosion rate of silicate dust is assumed to be always three times that of carbon dust, i.e., $Y_{\rm Si} = 3Y_{\rm C}$.
		The initial gas density is $n_{\rm H,0} = 0.1 {\rm cm^{-3}}$,
		and the vertical grey dotted line indicates the cooling time $t_c = 156$ kyr.
	}
	\label{fig:effvstime}
\end{figure}

Thermal sputtering destroys small grains more efficiently, as small grains have a larger total cross-section for a given mass
(which leads to the inverse $a$-dependence in Eq. \ref{eq:sputeqn}).
In contrast,
large grains are destroyed primarily by nonthermal sputtering.
This is due to the extra $a$-dependence in Eq. \ref{eq:sputeqn} through $Y_{\rm nth}$, which tends to outweigh the inverse $a$-dependence (while $Y_{\rm th}$, on the other hand, is independent of $a$).
Physically, this is because small grains experience a stronger drag force than large grains,
leading to lower ${\rm v_{rel}}$ (and hence lower $Y_{\rm nth}$).
It is worth noting that
thermal sputtering only operates during the Sedov-Taylor phase and is terminated once the radiative cooling kicks in at $t=t_c$
as the temperature of the shell becomes too low for thermal sputtering to be effective.
On the other hand,
betatron acceleration operates most efficiently right after $t_c$ when the shell is further compressed as it loses thermal presure support.

In the upper panel of Fig. \ref{fig:effvstime},
we show the time evolution of the sputtering efficiencies $\epsilon_{\rm sp,C}$ and $\epsilon_{\rm sp,Si}$, respectively.
The sputtering efficiencies first increase, then reach a maximum value at $t\sim t_c$ {(which we define as the \textit{total} sputtering efficiency)}, 
and afterwards gradually decline as sputtering becomes inefficient but the shell still keeps on sweeping up the ISM mass.
Note that the difference between $\epsilon_{\rm sp,Si}$ and $\epsilon_{\rm sp,C}$ gradually decreases with time.
To understand the origin of this,
we show in the lower panel of Fig. \ref{fig:effvstime} the ratio of $M_{\rm sp,Si}$ to $M_{\rm sp,C}$ as a function of time.
Initially, $M_{\rm sp,Si} / M_{\rm sp,C} \approx 3$ which reflects the fact that the erosion rate of silicate dust is about three times higher than that of carbon dust (cf. Fig. \ref{fig:sputY}).
However,
as time evolves,
$M_{\rm sp,Si} / M_{\rm sp,C}$ gradually decreases and eventually settles to a value of 1.5.
The decreasing ratio can be partially explained by the fact that
$Y_{\rm Si} / Y_{\rm C}$ also decreases as the shock weakens and both $T$ and ${\rm v_{rel}}$ decrease.
However, there is another factor, which we demonstrate by running another simulation where we set $Y_{\rm Si}$ to be exactly $3Y_{\rm C}$.
In this model,
$M_{\rm sp,Si} / M_{\rm sp,C}$ still shows a decline with time with a terminal value of 2
(which obviously cannot be explained by $Y_{\rm Si}/Y_{\rm C}$, which is constant).
This decline simply reflects the fact that the sputtered mass becomes limited by the available dust mass in the shocked gas.
Indeed,
in the limit where all of the dust in the shocked gas is destroyed,
we are bound to end up with $M_{\rm sp,Si} = M_{\rm sp,C}$ even if silicate dust is sputtered about three times faster than carbon dust.

\subsection{Final sputtered mass}

\subsubsection{The grain-size dependence}

\begin{figure*}
	\centering
	\includegraphics[trim = 10mm 15mm 10mm 0mm, clip, width=0.99\linewidth]{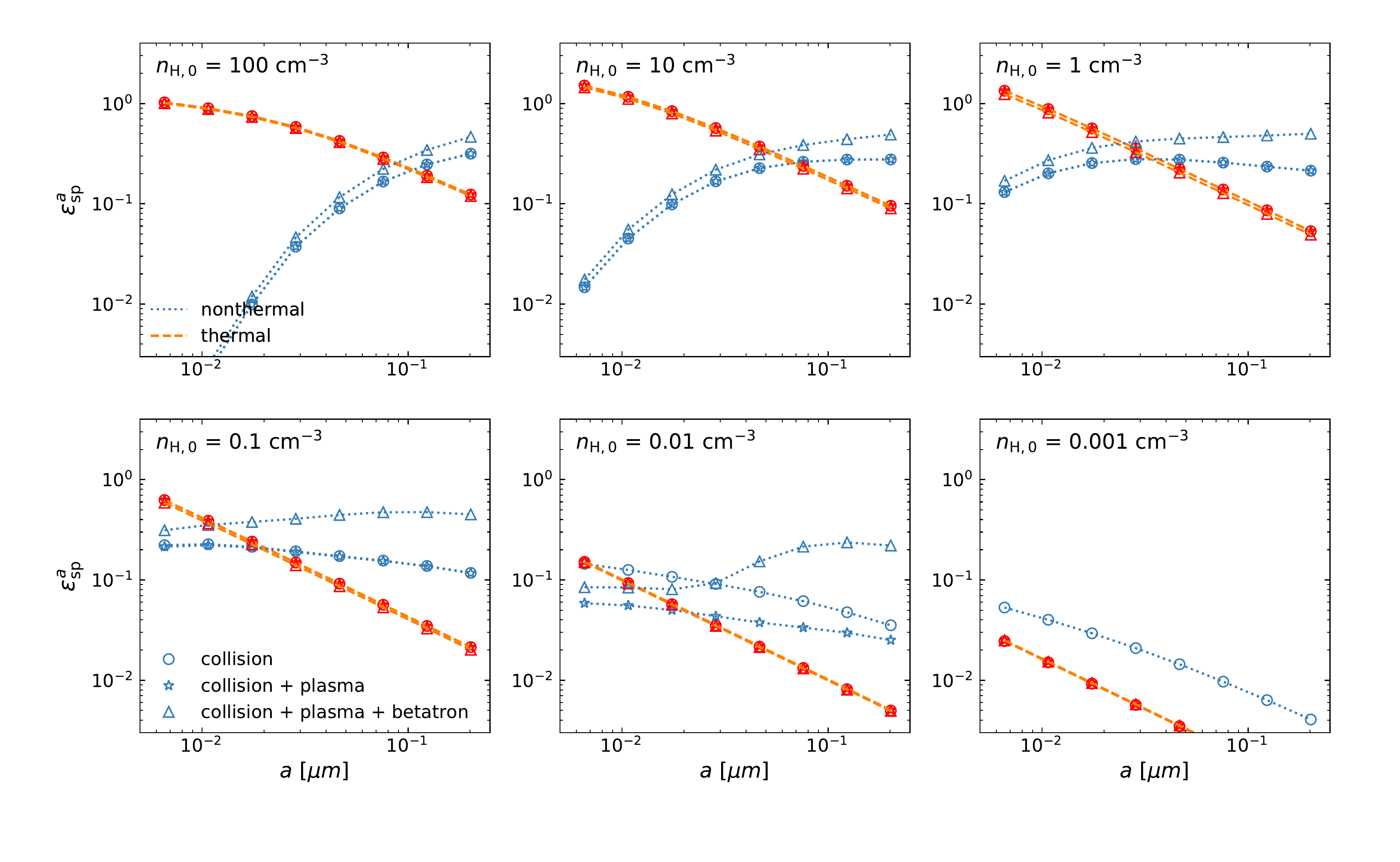}
	\caption{
		The relationship between grain size $a$ and the {total sputtering efficiency} in each $a$-bin for the silicate dust {$\epsilon^{a}_{\rm sp,Si}$} by both thermal (orange dashed) and nonthermal (blue dotted) sputtering, respectively.
		The panels show different values of the initial ambient density $n_{\rm H,0}$.
		Three different cases for the dust dynamics are compared in each panel:
		(i) direct collision only (circles), (ii) direct collision + plasma drag (stars) and (iii) direct collision + plasma drag + betatron acceleration (triangles, our fiducial model).
		{The efficiency for thermal sputtering is inversely proportional to $a$. 
		On the other hand,
		the relationship between the nonthermal sputtering efficiency and $a$ is more complicated and is density dependent.}
	}
	\label{fig:fig3finalmsp}
\end{figure*}

In Fig. \ref{fig:fig3finalmsp},
we show the relationship between $a$ and the {total sputtered efficiency} in each $a$-bin for the silicate dust {$\epsilon^{a}_{\rm sp,Si}$} by both thermal
and nonthermal
sputtering, respectively.
The initial ambient density $n_{\rm H,0}$ is systematically varied across different panels.
Three different physical models for the dust dynamics (which only affects nonthermal sputtering) are compared in each panel:
(i) direct collision only, (ii) direct collision + plasma drag and (iii) direct collision + plasma drag + betatron acceleration (our fiducial model).

For thermal sputtering,
there is an inverse correlation between $\epsilon^{a}_{\rm sp,Si}$ and $a$ which scales as {$\epsilon^{a}_{\rm sp,Si} \propto a^{-1}$
which is expected from Eq. \ref{eq:sputeqn}.}
The $a^{-1}$-scaling starts to break down in dense environments ($n_{\rm H,0} > 10 {\rm cm ^{-3}}$) as sputtering becomes limited by the available dust mass in the shocked gas.
This happens first to the smallest grains which have the highest sputtering rate,
causing the curve to {flatten out close to 1.
Note that $\epsilon^{a}_{\rm sp,Si}$ can slightly exceed unity as we enforce a fixed size distribution which effectively transfers mass from large grains to small grains.
}

The nonthermal sputtering is more complicated due to the extra $a$-dependence in $Y_{\rm nth}$ (while $Y_{\rm th}$ is independent of $a$).
Large grains experience weaker drag force and would typically reach higher ${\rm v_{rel}}$
{and thus higher $Y_{\rm nth}$, countering the $a^{-1}$-scaling.
In fact,
in the absence of cooling and betatron acceleration,
the $a$-dependence in the sputtering rate and drag force would cancel out exactly
and $\epsilon^{a}_{\rm sp,Si}$ would become independent of $a$.
Cooling breaks this exact cancellation as small grains decelerate faster and experience a different drag force due to its temperature dependence.
Betatron acceleration also breaks the cancellation as it is not explicitly $a$-dependent.
In a typical SN environment ($n_{\rm H,0} = 0.1 {\rm cm ^{-3}}$),
$\epsilon^{a}_{\rm sp,Si}$ ends up being insensitive to $a$.
This is an important property as it implies that the results should not be sensitive to the assumed size distribution.
}
There is a transition value of $a$ below which thermal sputtering dominates and above which nonthermal sputtering dominates.
This transition point generally increases with $n_{\rm H,0}$,
as the drag force is stronger in higher density gas, which leads to less efficient nonthermal sputtering.

Plasma drag has a negligible effect compared to direct collisions at $n_{\rm H,0} \gtrsim 0.1 {\rm cm ^{-3}}$,
but it becomes quite important in suppressing $\epsilon^{a}_{\rm sp,Si}$ at $n_{\rm H,0} < 0.1 {\rm cm ^{-3}}$ as the grain charge increases.
On the other hand,
betatron acceleration provides an efficient mechanism to enhance $\epsilon^{a}_{\rm sp,Si}$ especially at $n_{\rm H,0} = 0.1  {\rm cm ^{-3}}$,
which happens to be the typical density where SNe occur for solar-neighborhood conditions \citep{2017MNRAS.466.1903G, 2017MNRAS.466.3293P, 2017ApJ...846..133K}.


\begin{figure*}
	\centering
	\includegraphics[trim = 20mm 5mm 20mm 10mm, clip, width=0.99\linewidth]{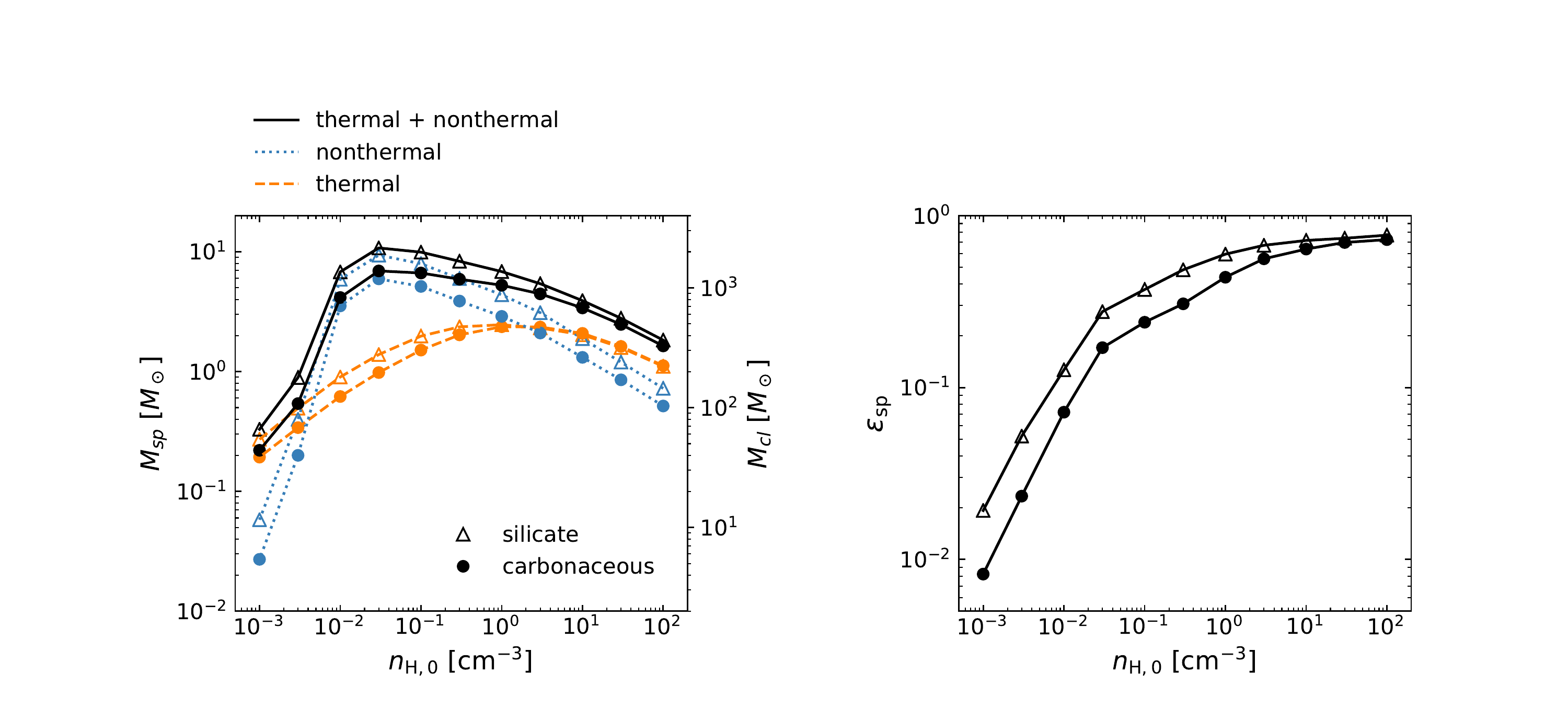}
	\caption{
		\textit{Left panel}:
		grain-size integrated final sputtered mass for the silicate dust $M_{\rm sp,Si}$ (triangles) and the carbon dust $M_{\rm sp,C}$ (circles)
		as a function of $n_{\rm H,0}$.
		Contributions from the thermal and nonthermal sputtering are shown as the orange dashed lines and the
		blue dotted lines, respectively,
		while the total (thermal + nonthermal) $M_{\rm sp}$ are shown as the black solid lines.
		The gas mass cleared of dust $M_{\rm cl}$ is shown in the right axis,
		with fitting formulae provided in Eq. \ref{eq:fitting}.
		\textit{Right panel}:
		total sputtering efficiency as a function of $n_{\rm H,0}$.
		}
	\label{fig:fig4finalmspvsnh}
\end{figure*}

\subsubsection{Grain-size integrated final sputtered mass}
	
In the left panel of Fig. \ref{fig:fig4finalmspvsnh},
we show the final $M_{\rm sp,Si}$
and $M_{\rm sp,C}$
as a function of $n_{\rm H,0}$.
Contributions from the thermal and nonthermal sputtering are shown, along with the total (thermal + nonthermal) sputtered mass.
The gas mass cleared of dust is shown in the right axis (recall that $M_{\rm cl} = M_{\rm sp}/ D_0$).

As $n_{\rm H,0}$ increases,
thermal sputtering becomes more efficient (cf. Eq. \ref{eq:sputeqn}) while $M_{\rm swept} \propto n_{\rm H,0}^{-0.26}$ decreases,
making $M^{\rm th}_{\rm sp}$ a concave function of $n_{\rm H,0}$,
peaking at $n_{\rm H,0} \approx 1 {\rm cm^{-3}}$.
On the other hand,
since nonthermal sputtering depends sensitively on the drag force which is stronger in denser gas,
the peak of $M_{\rm sp,nth}$ is shifted towards a lower value at $n_{\rm H,0} \approx 0.03 {\rm cm^{-3}}$,
below which the plasma drag starts to become very efficient in slowing down the dust, causing the downward bending.
Nonthermal sputtering dominates over thermal sputtering in the range of $0.003 < n_{\rm H,0} < 10 {\rm cm^{-3}}$,
which covers the range where most SNe occur.
The right panel of Fig. \ref{fig:fig4finalmspvsnh} shows the total sputtering efficiency $\epsilon_{\rm sp,C}$ and $\epsilon_{\rm sp,Si}$ (which occurs at $t \sim t_c$) as a function of $n_{\rm H,0}$.

The efficiency monotonically increases with $n_{\rm H,0}$,
indicating that the decrease of $M_{\rm sp}$ in the left panel at high $n_{\rm H,0}$  is indeed due to a decrease of $M_{\rm swept}$.
It plateaus at $n_{\rm H,0} \gtrsim 10 {\rm cm^{-3}}$ as sputtering becomes limited by the available dust.
For the same reason,
the ratio of $\epsilon_{\rm sp,Si}$ to $\epsilon_{\rm sp,C}$ is a monotonically decreasing  function of $n_{\rm H,0}$, from a factor of 2.5 in the diffuse gas (due to the difference in the erosion rates) to almost unity in the dense gas.

\begin{table}
	\centering
	\begin{tabular}{c | c | c | c | c | c | c}
		\hline	
		\hline	
		 $n_{\rm H,0}$ & $M^{\rm th}_{\rm cl,C} $ & $M^{\rm nth}_{\rm cl,C}$ & $M_{\rm cl,C}$ & $M^{\rm th}_{\rm cl,Si} $ &  $M^{\rm nth}_{\rm cl,Si}$ & $M_{\rm cl,Si}$\\
		\hline	
		 0.001  & 38.6    &   5.41    &   44.0    & 54.3   &   11.6 &   65.9\\
		 0.003 & 68.0    &   40.0    &   108     & 98.8   &  79.7  &   179\\
		 0.01    & 124      &   706     &   830     & 180     &  1180 &   1360\\
		 0.03   & 196       &   1190   &   1390   &  277    &  1880 &   2157\\
		 0.1      & 302      &   1030   &   1330   &  395   &  1590 &   1990\\
		 0.3     & 405      &   776     &   1180    & 472     &   1190&   1660\\
		 1         & 473       &   578    &   1050   & 492     &  874  &    1370\\
		 3        & 471        &   420    &   891     & 461     &   620  &   1080\\
		 10       & 417       &   263    &   680     & 405     &   376 &    780\\
		 30      & 324      &   171      &   495     &  317     &  240  &    557\\
		 100     & 224      &   103     &   327     &  220   &  145   &    365\\
		 \hline	
	\end{tabular}
	\caption{
		Gas mass cleared of dust ($M_{\rm cl}$) for different values of initial density $n_{\rm H,0}$. 
	}\label{tbl:Mclear}
\end{table}

For a typical SN environment $n_{\rm H,0} = 0.3 {\rm cm}^{-3}$ 
and assuming $D_{\rm C,0} = D_{\rm Si,0} = 0.005$,
a supernova destroys 5.9 M$_\odot$ of carbon dust and 8.3 M$_\odot$ of silicate dust,
corresponding to $M_{\rm cl,C} = $ 1180 M$_\odot$ and $M_{\rm cl,Si} = $ 1660 M$_\odot$, respectively.
The results for the mass of dust cleared for different initial gas densities are summarized in Table \ref{tbl:Mclear}.
We also provide simple fitting formulae for the gas mass cleared of dust below:
\begin{eqnarray}\label{eq:fitting}
y^{\rm nth}_{\rm Si} &=& 2.93 - 0.37x + 0.010x^2 + 0.040x^3 - 0.024x^4, \nonumber\\
y^{\rm nth}_{\rm C} &=& 2.74 - 0.36x + 0.023x^2 + 0.049x^3 - 0.028x^4, \nonumber\\
y^{\rm th}_{\rm Si} &=& 2.70 + 0.018x - 0.10x^2, \nonumber\\
y^{\rm th}_{\rm C} &=& 2.65 + 0.061x - 0.10x^2, 
\end{eqnarray}	
where 
$y^{\rm th}_{\rm Si}\equiv \log_{10} (M^{\rm th}_{\rm cl,Si} / {\rm M_\odot})$,
$y^{\rm th}_{\rm C}\equiv \log_{10} (M^{\rm th}_{\rm cl,C} / {\rm M_\odot})$,
$y^{\rm nth}_{\rm Si}\equiv \log_{10} (M^{\rm nth}_{\rm cl,Si} / {\rm M_\odot})$,
$y^{\rm nth}_{\rm C}\equiv \log_{10} (M^{\rm nth}_{\rm cl,C} / {\rm M_\odot})$,
and $x\equiv \log_{10} (n_{\rm H,0} / {\rm cm^{-3}})$.

\begin{figure}
	\centering
	\includegraphics[trim = 10mm 0mm 10mm 10mm, clip, width=0.9\linewidth]{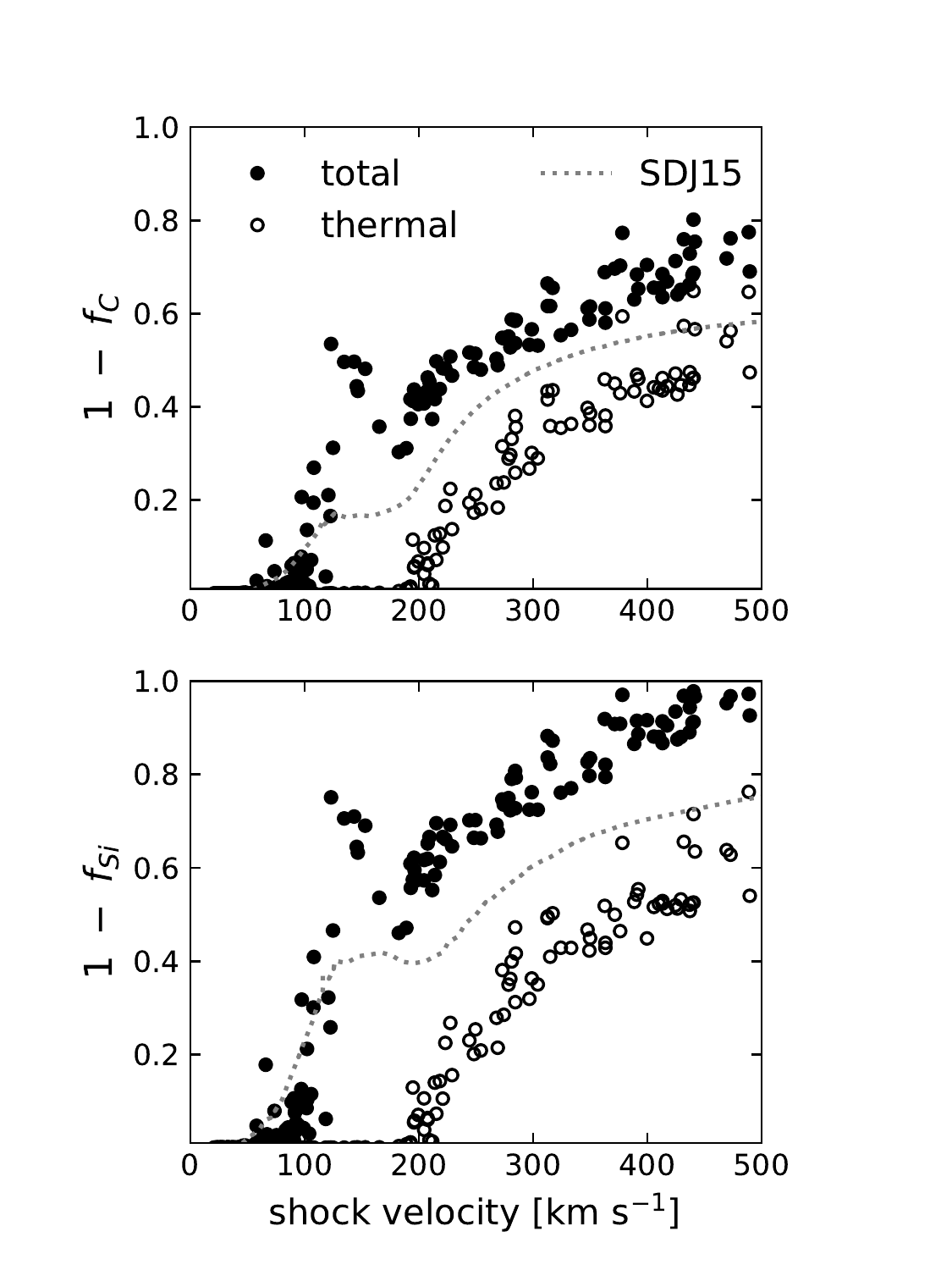}
	\caption{
		The final sputtered mass fraction of each gas particle ($1 - f$) vs. the shock velocity $v_{\rm shock}$ for $n_{\rm H,0} = 0.3 {\rm cm}^{-3}$. 
		The upper and lower panels are for the carbon and silicate dust, respectively.
		The empty circles represent the thermal sputtering while the filled circles represent the total sputtering (thermal + nonthermal).
		The dotted lines are the results of SDJ15.
		In general, higher $v_{\rm shock}$ leads to more sputtering,
		while the bump at $v_{\rm shock}\sim 150 {\rm km/s}$ is due to betatron acceleration.
	}
	\label{fig:shockvelvseff}
\end{figure}

In Fig. \ref{fig:shockvelvseff},
we show the final sputtered mass fraction of each gas particle ($1 - f$) vs. the shock velocity $v_{\rm shock}$ for $n_{\rm H,0} = 0.3 {\rm cm}^{-3}$. 
We measure $v_{\rm shock}$ by recording the maximum radial gas velocity of a particle during the entire simulation (which gives the post-shock gas velocity) multiplied by a factor of $4/3$ (as the shock velocity is $3/4$ times the post-shock gas velocity).
The upper and lower panels are for the carbon and silicate dust, respectively.
The empty circles represent the thermal sputtering while the filled circles represent the total sputtering (thermal + nonthermal).
For comparison, we also show the results of SDJ15 for the total sputtered mass fraction.
Gas located at a smaller radius will experience a higher $v_{\rm shock}$.
In general,
$1-f$ increases with $v_{\rm shock}$,
as both $T$ and ${\rm v_{rel}}$ increase with $v_{\rm shock}$.
Thermal sputtering is efficient only for $v_{\rm shock} > 200 {\rm km/s}$,
below which the SNR has already entered the radiative cooling phase and hence the shell becomes too cold for thermal sputtering to operate.
On the other hand,
nonthermal sputtering remains efficient for $v_{\rm shock} < 200 {\rm km/s}$,
and even exhibits a local maximum at $v_{\rm shock}\approx 150 {\rm km/s}$.
This comes from the contribution of betatron acceleration,
which becomes most efficient right after $t_c$ as the shell is compressed.
Sputtering becomes negligible when $v_{\rm shock} < 100 {\rm km/s}$.
{Our $1-f$ is around $10-20\%$ higher than what is found in SDJ15.
The origin of this discrepancy is not exactly clear,
but the potential candidates are
(i) the magnetic pressure included in SDJ15 (effectively) but not in our model and 
(ii) the difference in the grain-size distribution which we assumed to be constant while SDJ15 follows the size evolution\footnote{We have verified that our results are not sensitive to the adopted SN energy by carrying out an additional simulation with $E_{51} = 0.5$.}.
}
However,
the same trend can be observed in their calculation,
where $1-f$ decreases with $v_{\rm shock}$ but there is a small bump at $v_{\rm shock} \sim 170{\rm km/s}$ (less pronounced compared to ours),
which may also be due to betatron acceleration.

\begin{figure*}
	\centering
	\includegraphics[trim = 5mm 5mm 5mm 5mm, clip, width=0.97\linewidth]{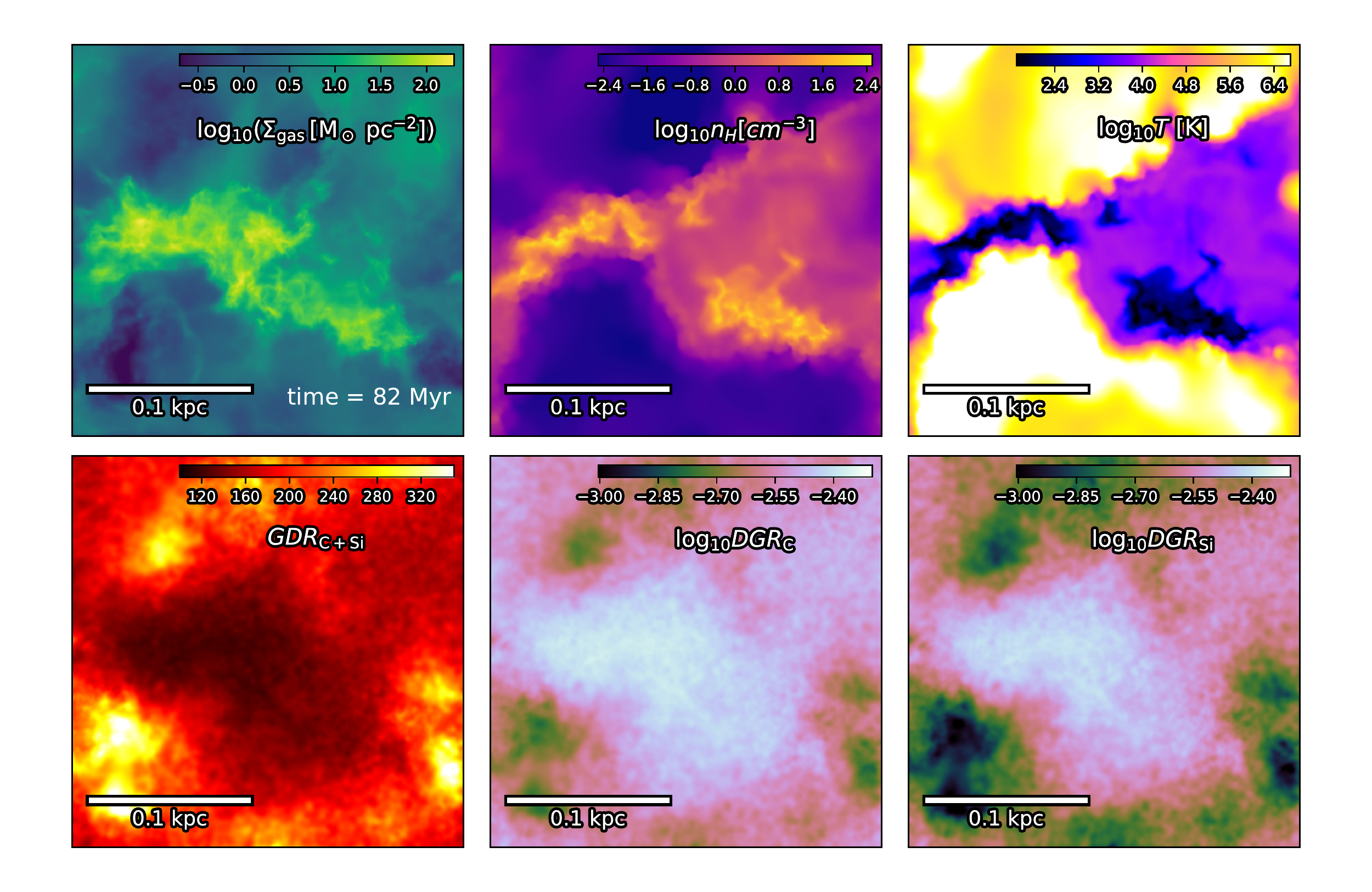}
	\caption{A turbulent multiphase ISM which resembles the solar-neighborhood conditions driven by randomly (spatially uniform) injected SNe with a prescribed rate ($R_{\rm SN} = 2{\rm Myr^{-1}}$) at $t = 82.3$ Myr.
		\textit{Upper left}: gas surface density;
		\textit{upper middle}: gas temperature (slice);
		\textit{upper right}: gas hydrogen number density (slice).
		\textit{lower left}: projected total (carbon+silicate) gas-to-dust ratio (GDR);
		\textit{lower middle}: projected DGR for carbon dust;
		\textit{lower right}: projected DGR for silicate dust.
	}
	\label{fig:dgrmap}
\end{figure*}

\section{SN-driven multiphase ISM}\label{sec:SNbox}

\begin{figure}
	\centering
	\includegraphics[trim = 5mm 20mm 0mm 5mm, clip, width=0.9\linewidth]{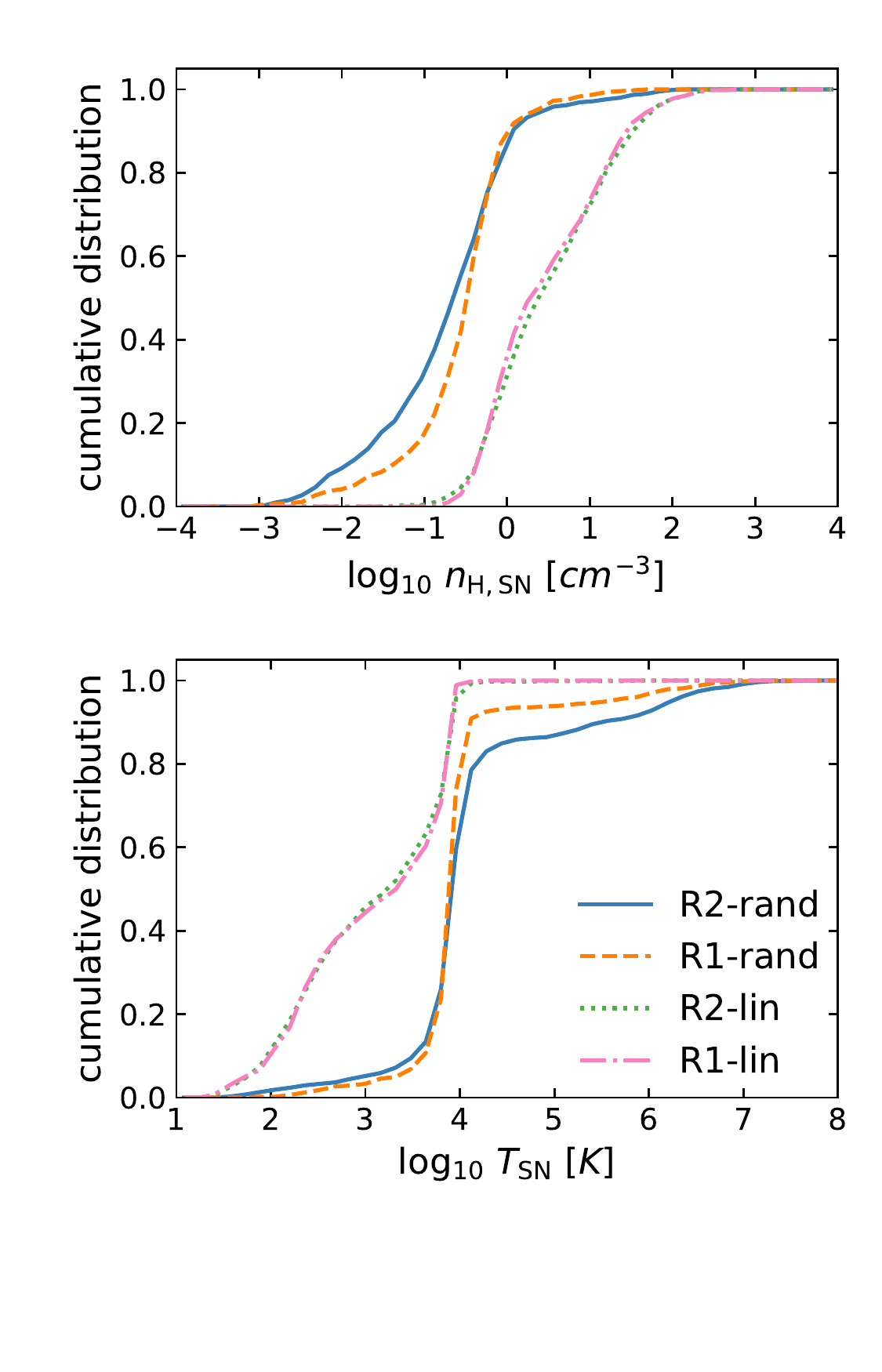}
	\caption{
		Fraction of SNe that occur at $n_{\rm H} < n_{\rm H,SN}$ (left) and $T < T_{\rm SN}$ (right).
		With random driving,
		SNe occur mostly in the diffuse warm gas ($n_{\rm H} \sim 0.2 {\rm cm^{-3}}$ and $T\sim 10^4$K) and 10$-$20\% occur in pre-existing hot SN bubbles.
		With linear driving,
		SNe occur in a denser and cooler environment, and almost never in pre-existing hot bubbles.
	}
	\label{fig:fig9snenv}
\end{figure}

In this section,
we investigate dust sputtering in a multiphase ISM that resembles the solar-neighborhood environment. 
The initial conditions are similar to those in the single SN case.
We set up a uniform and static medium with $n_{\rm H,0} = 1 {\rm cm^{-3}}$ and $T_0 = 10^4$K in a cubic box.
The box is 0.25 kpc on a side with periodic boundary conditions and thus the total gas mass in the box is $M_{\rm gas} = 5\times10^5 {\rm M_\odot}$.
{The periodic boundary conditions are appropriate for galaxies that have little galactic outflows, which applies to our case.}
The gas particle mass is $m_{\rm gas} = 0.5 {\rm M_\odot}$.
The initial total DGR is 0.01 which is consistent with observations (e.g. \citealp{2003ApJ...582..846K}),
and we assume equal partition for the two dust species, i.e., $D_{\rm C} = D_{\rm Si} = 0.005$\footnote{
We note that the choice of the initial DGR does not affect our main results such as dust destruction timescales and DGR inhomogeneity.}.
SNe are injected stochastically into the ISM with a prescribed rate $R_{\rm SN}$, driving turbulence and thermal instability.
{Self-gravity is not included, and therefore we expect that the collapse of the dense clouds will not be realistically captured.
However, most of the SNe occur in the more diffuse phase that can be properly modeled with our setup,
which has been shown to be capable of creating a realistic multiphase ISM structure \citep{2015MNRAS.449.1057G, 2015ApJ...814....4L, 2016ApJ...822...11P}.}
The stochastic injection of SNe can be either be density-independent, or have a correlation with local gas density,
and is implemented as follows.
Each gas particle has a probability $p_{\rm SN}$ to be chosen as a location to inject a SN:
\begin{equation}
	p_{\rm SN} = \frac{R_{\rm SN} \Delta t }{N_{\rm gas}} \delta^{\alpha - 1},
\end{equation}
where $R_{\rm SN}$ is the target SN rate of the system,
$\Delta t$ is the timestep,
$N_{\rm gas}$ is the number of gas particles,
$\delta \equiv n_{\rm H} / n_{\rm H,0}$ is the local over-density,
and $\alpha$ is an index controlling the density dependence.
Since $p_{\rm SN}$ is a probability \textit{per particle}
and particles are intrinsically mass weighted (i.e. they cluster in high density regions by construction),
the SN rate \textit{per volume} will end up being proportional to $\delta^\alpha$.
Therefore,
$\alpha$ = 0 will result in SNe being injected randomly (uniformly distributed in space),
while $\alpha = 1$ will inject SNe preferentially in dense gas.
We will refer to the former as ``random driving'' and the latter as ``linear driving''.
Random driving is the preferred model as it leads to SNe occurring mostly in the diffuse gas in the range of $0.01 < n_{\rm H} < 1 {\rm cm}^{-3}$,
which is consistent with more sophisticated simulations that include self-gravity and star formation \citep{2017MNRAS.466.1903G, 2017MNRAS.466.3293P, 2017ApJ...846..133K}.
Linear driving is considered less realistic and serves as a comparison case.
Once the SN location is chosen,
the injection scheme is similar to that used in the single SN case,
except that we adopt $N_{\rm inj} = 32$ such that the mass of energy injection is still $N_{\rm inj} m_{\rm gas} = 16 {\rm M_\odot}$.
Assuming that there is one SN for every 100 M$_\odot$ of newly formed stars,
the SN rate is determined as $R_{\rm SN} = M_{\rm gas} / (100{\rm M_\odot} t_{\rm dep})$
where $t_{\rm dep} \sim 2.4$ Gyr as suggested by spatially resolved observations of nearby disk galaxies (e.g.  \citealp{2008AJ....136.2846B, 2011ApJ...730L..13B}).
This leads to $R_{\rm SN} = 2{\rm Myr^{-1}}$ as our fiducial choice
{corresponding to 2.68 SNe per 100 years in the Milky Way which has a total gas mass of $6.7\times 10^9 {\rm M_\odot}$ \citep{2011piim.book.....D}.}
We also explore a lower SN rate ($R_{\rm SN} = 1{\rm Myr^{-1}}$) for comparison.
In summary, we investigate four cases:
\begin{enumerate}
	\item \textit{R2-rand}: $R_{\rm SN} = 2 {\rm Myr^{-1}}$ with random driving (fiducial model).
	\item \textit{R1-rand}: $R_{\rm SN} = 1 {\rm Myr^{-1}}$ with random driving.
	\item \textit{R2-lin}: $R_{\rm SN} = 2 {\rm Myr^{-1}}$ with linear driving.
	\item \textit{R1-lin}: $R_{\rm SN} = 1 {\rm Myr^{-1}}$ with linear driving.
\end{enumerate}
Each simulation is run for 0.4 Gyr.

In Fig. \ref{fig:dgrmap},
we show the maps of gas surface density (upper left), $n_{\rm H}$ (upper middle, slice), $T$ (upper right, slice), 
projected total (carbon+silicate) gas-to-dust ratio (GDR, lower left), projected DGR for carbon dust (lower middle) and projected DGR for silicate dust (lower right) for the \textit{R2-rand} model at $t = 82.3$ Myr.
The projected DGR is calculated as the ratio of the dust surface density to the gas surface density,
while the GDR is its inverse which is more frequently shown in observational studies.
The ISM is turbulent, structured and multiphase.
SNe create hot ($T\gtrsim10^6$K) and diffuse regions where the DGR is significantly lower than average due to sputtering.
Between SN events, however,
the DGR can be quickly homogenized by turbulent mixing.
Note that there is no sub-grid turbulent mixing between particles in our calculation and so the mixing is purely due to particle motions.
Silicate dust is more inhomogeneous than carbon dust due to its higher erosion rate,
especially in the SN bubbles.

\begin{figure}
	\centering
	\includegraphics[trim = 5mm 20mm 0mm 12mm, clip, width=0.99\linewidth]{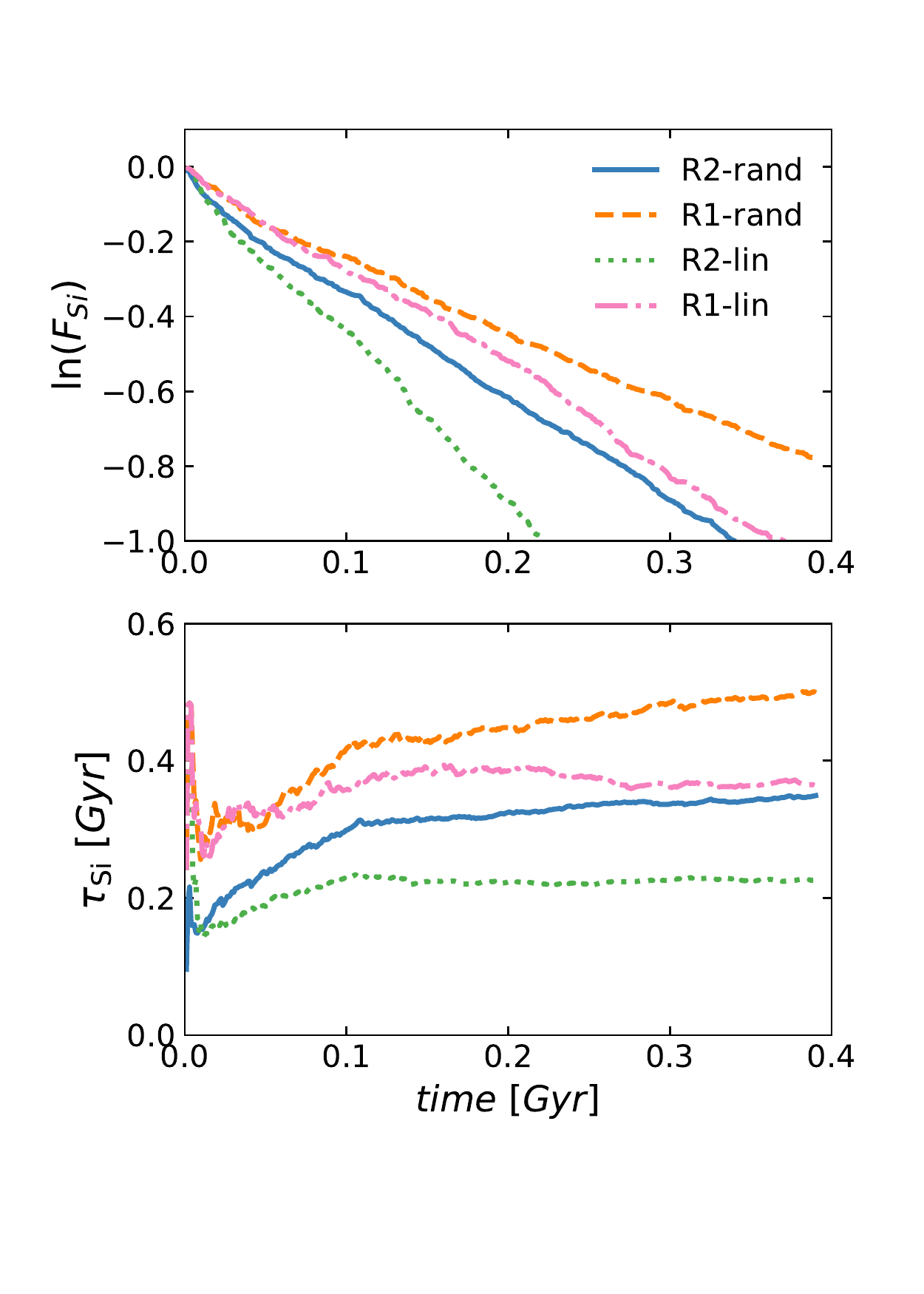}
	\caption{
		Time evolution of the silicate dust mass fraction $F_{\rm Si}$ (upper panel) and the destruction timescale for silicate dust $\tau_{\rm Si}\equiv  {-t} / {\ln F_{\rm Si}}$ (lower panel) in a multiphase ISM which resembles the solar-neighborhood conditions. 
		SNe are injected into the ISM with a prescribed rate $R_{\rm SN}$ and with a controlled density dependence (random driving and linear driving).
		Our fiducial model is random driving with $R_{\rm SN} = 2 {\rm Myr^{-1}}$ (\textit{R2-rand})
		where $\tau_{\rm Si} = 0.35$Gyr.
		With the same $R_{\rm SN}$,
		random driving leads to a longer $\tau_{\rm Si}$ compared to linear driving due to SN clustering,
		which leads to more SNe occurring in pre-existing bubbles where there is little dust left to be destroyed.		
	}
	\label{fig:tSN_estimate}
\end{figure}

In Fig. \ref{fig:fig9snenv},
we show the fraction of SNe that occur at $n_{\rm H} < n_{\rm H,SN}$ (left) and $T < T_{\rm SN}$ (right).
With random driving,
SNe occur mostly in the diffuse warm gas ($n_{\rm H,SN} \sim 0.2 {\rm cm^{-3}}$ and $T_{\rm SN}\sim 10^4$K). 
Around 10$-$20\% of SNe occur where $T_{\rm SN} > 10^4$K,
which is indicative of pre-existing SN bubbles as no other heating mechanisms can heat the gas above $10^4$K.
As the SN locations are randomly chosen,
the distributions of $n_{\rm H,SN}$ and $T_{\rm SN}$ just reflect the volume-weighted distributions of the gas conditions ($n_{\rm H}$ and $T$).
Therefore,
the volume filling fraction of hot gas ($T_{\rm SN} > 10^4$K) is subdominant (10\% for \textit{R1-rand} and 20\% for \textit{R2-rand}, respectively)
and most of the volume is filled by the warm gas.
This is in very good agreement with \citet{1998ApJ...503..700F} who deduced that the hot gas volume filling factor in the solar neighborhood is $\lesssim$ 20\% based on observational data and analytic calculations.
In contrast,
with linear driving,
SNe occur in denser environments in the range $1 < n_{\rm H} < 30 {\rm cm}^{-3}$,
leading to smaller SN bubbles which fade away faster (due to the shorter cooling times), and therefore to a more homogeneous ISM.
Due to the combined effect of the density dependence of $p_{\rm SN}$ per se and the resulting ISM structure,
SNe rarely occur in pre-existing hot bubbles.

\subsection{Dust destruction timescale}

\begin{table}
	\centering
	\begin{tabular}{c | c | c | c | c}
		& \textit{R2-rand} & \textit{R1-rand} & \textit{R2-lin} & \textit{R1-lin} \\
		&(fiducial model)&\\
		\hline	
		\hline	
		$\tau_{\rm Si}$ (Gyr) & 0.35 & 0.50 & 0.23 & 0.36\\
		\hline	
		$\tau_{\rm C}$ (Gyr) & 0.44 & 0.64 & 0.29 & 0.46\\
		\hline
	\end{tabular}
	\caption{
		Dust destruction timescales for the carbon dust ($\tau_{\rm C}$) and silicate dust ($\tau_{\rm Si}$) in four different models.
		\textit{R2-rand} is our fiducial model.
	}\label{tbl:tdust}
\end{table}

The dust destruction timescale is perhaps the most useful summary statistic that can be obtained from modeling dust sputtering in the multiphase ISM.
We will focus the discussion on the silicate dust as the carbon dust behaves in a qualitatively similar manner.
Given the total mass of the silicate dust $M_{\rm Si}$,
its time evolution can be described by $M_{\rm Si}(t) = M_{\rm Si}(0) \exp(-t / \tau_{\rm Si} )$ where $\tau_{\rm Si}$ is the dust destruction timescale.
Therefore, we can calculate
\begin{equation}
\tau_{\rm Si} = \frac{-t}{\ln [M_{\rm Si}(t) / M_{\rm Si}(0)]} = \frac{-t}{\ln F_{\rm Si}(t)}
\end{equation}
at any given $t$.
In Fig. \ref{fig:tSN_estimate},
we show the time evolution of $F_{\rm Si}$ (upper panel) and $\tau_{\rm Si}$ (lower panel).
The destruction timescales at the end of the simulations ($t=0.4$Gyr) for both carbon and silicate dust are shown in Table \ref{tbl:tdust}.

The system settles into a quasi-steady state in the sense that $\tau_{\rm Si}$ is almost constant in time after 0.1 Gyr.
Obviously,
$R_{\rm SN}$ should be an important parameter which explicitly controls $\tau_{\rm Si}$.
However, we find that the dependence of $\tau_{\rm Si}$ on $R_{\rm SN}$ is sub-linear: decreasing $R_{\rm SN}$ by a factor of two only increases $\tau_{\rm Si}$ by a factor of 1.5.
This is likely due to the dependence of ISM structure on $R_{\rm SN}$ which also affects $\tau_{\rm Si}$.
Interestingly, 
the SN environment also has a significant effect on $\tau_{\rm Si}$.
With the same $R_{\rm SN}$,
random driving leads to a longer $\tau_{\rm Si}$ compared to linear driving.
This may seem counterintuitive as we have seen that the sputtered mass per SN is slightly higher in more diffuse environments (cf. Fig. \ref{fig:fig4finalmspvsnh}).
However,
with random driving,
SNe occur more frequently in pre-existing low-DGR SN bubbles and therefore sputtering becomes limited by the available dust.
In an extreme case where all dust has already been destroyed in the SN bubbles,
the subsequent SNe will become ``futile events'' in terms of dust destruction.
As such, when
SNe explode in pre-existing bubbles, the net effect on dust destruction is effectively similar to a situation with a lower SN rate,
as pointed out by \citet{1989IAUS..135..431M}.

The destruction timescale for the carbon dust is longer than that for the silicate dust by a factor of 1.25 in all of our four models.
With linear driving,
this is expected as $M_{\rm sp,Si}$ to $M_{\rm sp,C}$ per SN is also about 1.25 for $n_{\rm H,0} \sim 3 {\rm cm^{-3}}$ where the majority of SNe occur.
However,
with random driving,
most SNe occur in gas with density $n_{\rm H} \sim 0.1 {\rm cm^{-3}}$ where $M_{\rm sp,Si} / M_{\rm sp,C}$ per SN is about 1.5.
Namely,
the difference in $M_{\rm sp}$ between carbon dust and silicate dust does not translate to $\tau$ in a linear fashion.
The reason is related to the fact that SNe occur more frequently in pre-existing SN bubbles with random driving:
in these SN bubbles,
dust destruction is limited by the available dust and the difference in the sputtering rate becomes less important,
leading to the sub-linear scaling.

\begin{figure}
	\centering
	\includegraphics[trim = 5mm 10mm 5mm 20mm, clip, width=0.99\linewidth]{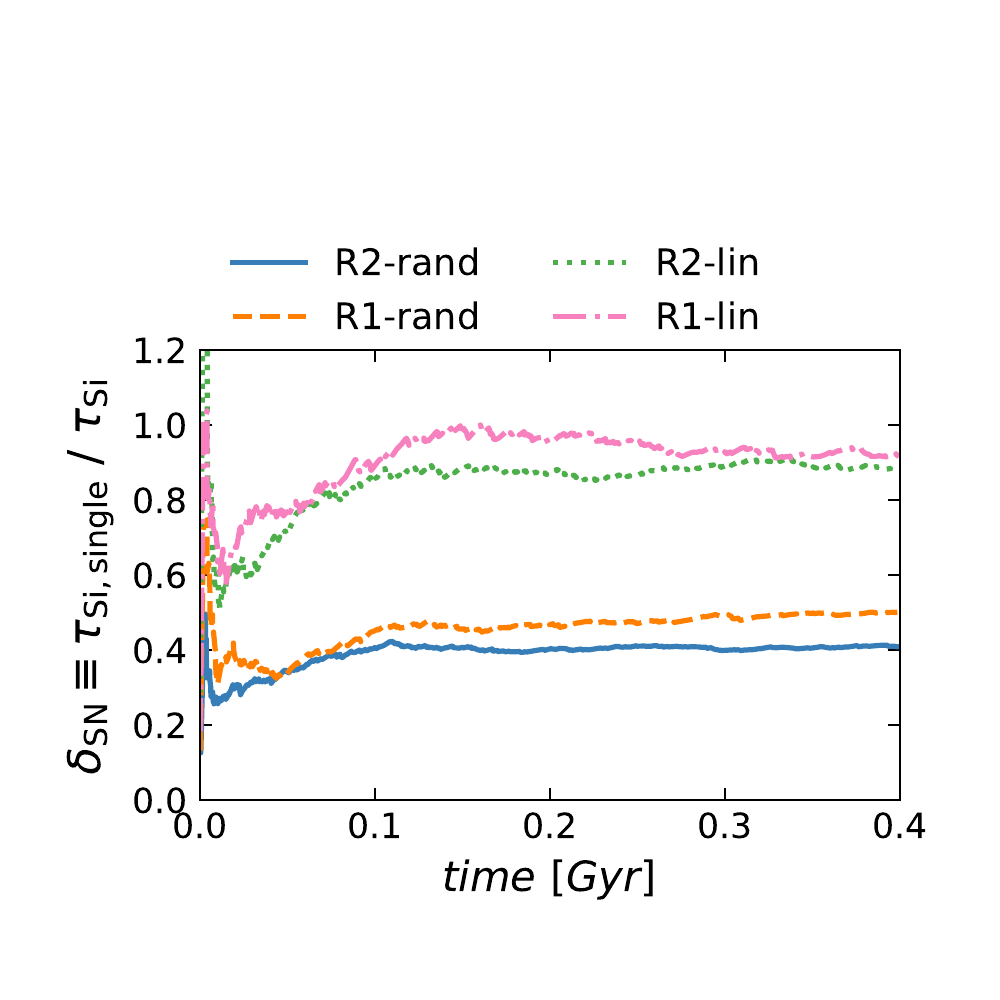}
	\caption{Time evolution of $\delta_{\rm SN} \equiv \tau_{\rm Si,single} / \tau_{\rm Si}$,
		where $\tau_{\rm Si,single}$ is the destruction timescale estimated from our single SN results using Eq. \ref{eq:fitting}.
		SNe that occur in low-DGR SN bubbles where there is not much dust left to be destroyed becomes ``futile events'' for dust destruction,
		which effectively increases the destruction timescales.
		As such, our results from single SNe underestimate the destruction timescales.
	}
	\label{fig:applysinglesn}
\end{figure}

It is interesting to see how our results from the single SN setup relate to the case of multiple SNe exploding in the multiphase ISM.
We calculate the amount of dust destroyed per SN based on its ambient density $n_{\rm H,SN}$ in the multiphase ISM using our fitting formulae Eq. \ref{eq:fitting}, 
i.e., $D_{\rm Si} M_{\rm cl,Si}(n_{\rm H,SN})$.
The total dust mass at time $t$ can therefore be estimated as
\begin{equation}
M_{\rm Si,single}(t) = M_{\rm Si,0} -  \sum_{i, t_i<t} \frac{M^i_{\rm Si,single}}{M_{\rm gas}} M_{\rm cl,Si}(n^i_{\rm H,SN}),
\end{equation}
where the index $i$ represents the $i$-th SN in the simulations and the summation is over all SN events with $t_i < t$.
The factor $M^i_{\rm Si,single} / M_{\rm gas}$ accounts for the decreasing DGR of the system (and hence the sputtered mass per SN) as time evolves.
The dust destruction timescale is then $\tau_{\rm Si,single} =  -t/ \ln (M_{\rm Si,single} / M_{\rm gas})=  -t/ \ln F_{\rm Si,single}$.
In Fig. \ref{fig:applysinglesn},
we show $\delta_{\rm SN} \equiv \tau_{\rm Si,single} / \tau_{\rm Si}$ as a function of time.
With linear driving (\textit{R2-lin} and \textit{R1-lin}),
the estimate based on the single SN results yields a dust destruction timescale which agrees with our full multiphase calculation without 10\% accuracy, which is quite good.
On the other hand,
with random driving (\textit{R2-rand} and \textit{R1-rand}),
our single SN results underestimate $\tau_{\rm Si}$ significantly.
This is, again, because there are SNe occurring in low-DGR SN bubbles where there is not much dust left to be destroyed.
As pointed out by \citet{1989IAUS..135..431M},
this effectively reduces the SN rate and increases $\tau_{\rm Si}$ by a factor of $\delta^{-1}_{\rm SN}$.
Our fiducial model (\textit{R2-rand}) predicts $\delta_{\rm SN} = 0.4$ which is in very good agreement with the estimate of \citet{1989IAUS..135..431M}, $\delta_{\rm SN} = 0.36$, based on observational data.
On the other hand,
our model \textit{R1-rand} predicts $\delta_{\rm SN} = 0.5$ which suggests that $\delta_{\rm SN}$ is not universal and depends on the SN rate.
It also explains why the relationship between $R_{\rm SN}$ and $\tau^{-1}_{\rm Si}$ is sub-linear as the SN clustering also changes with $R_{\rm SN}$.

\subsection{DGR inhomogeneity}

\begin{figure}
	\centering
	\includegraphics[trim = 2mm 5mm 5mm 10mm, clip, width=0.99\linewidth]{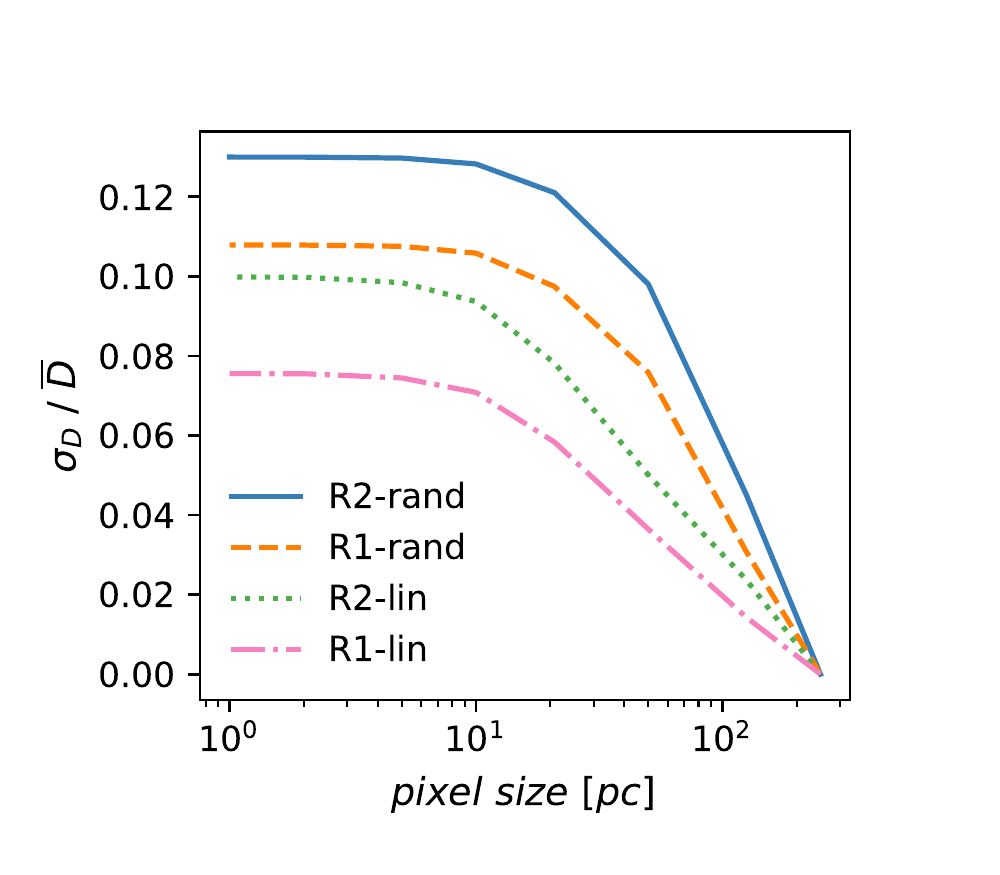}
	\caption{
		Time-averaged (from $t=$ 0.1 to 0.4 Gyr) standard score (normalized standard deviation) of DGR as a function of pixel size $\Delta x$.
		The DGR becomes more inhomogeneous when $R_{\rm SN}$ is higher and when SNe occur more frequently in the pre-existing SN bubbles (random driving).
	}
	\label{fig:stdvspixelsize}
\end{figure}

To quantify the spatial inhomogeneity of the ``total'' DGR $D = D_{\rm C} + D_{\rm Si}$,
we measure its standard score defined as
$z_{D} \equiv \sigma_{D} / \overline{D}$
where $\overline{D}$ and $\sigma_{D}$ are the spatial average and the standard deviation of DGR, respectively.
In Fig. \ref{fig:stdvspixelsize},
we show the time-averaged (from $t=0.1$ to 0.4 Gyr) $z_{\rm DGR}$
as a function of the ``pixel size'', 
which is the length-scale we use to smooth the projected DGR map with the SPH kernel function (cubic spline).
In all the four models, $z_{D}$ increases as $\Delta x$ and saturates at $\Delta x \lesssim 10$pc
at the level of 14\% (\textit{R2-rand}), 10\% (\textit{R1-rand}), 9\% (\textit{R2-lin}) and 6\% (\textit{R1-lin}), respectively.
The DGR is more inhomogeneous when $R_{\rm SN}$ is higher as SNe are the source of the inhomogeneity.
Interestingly,
the SN environment also plays a role: random driving results in a more inhomogeneous DGR than linear driving.

\begin{figure}
	\centering
	\includegraphics[trim = 5mm 0mm 18mm 10mm, clip, width=0.99\linewidth]{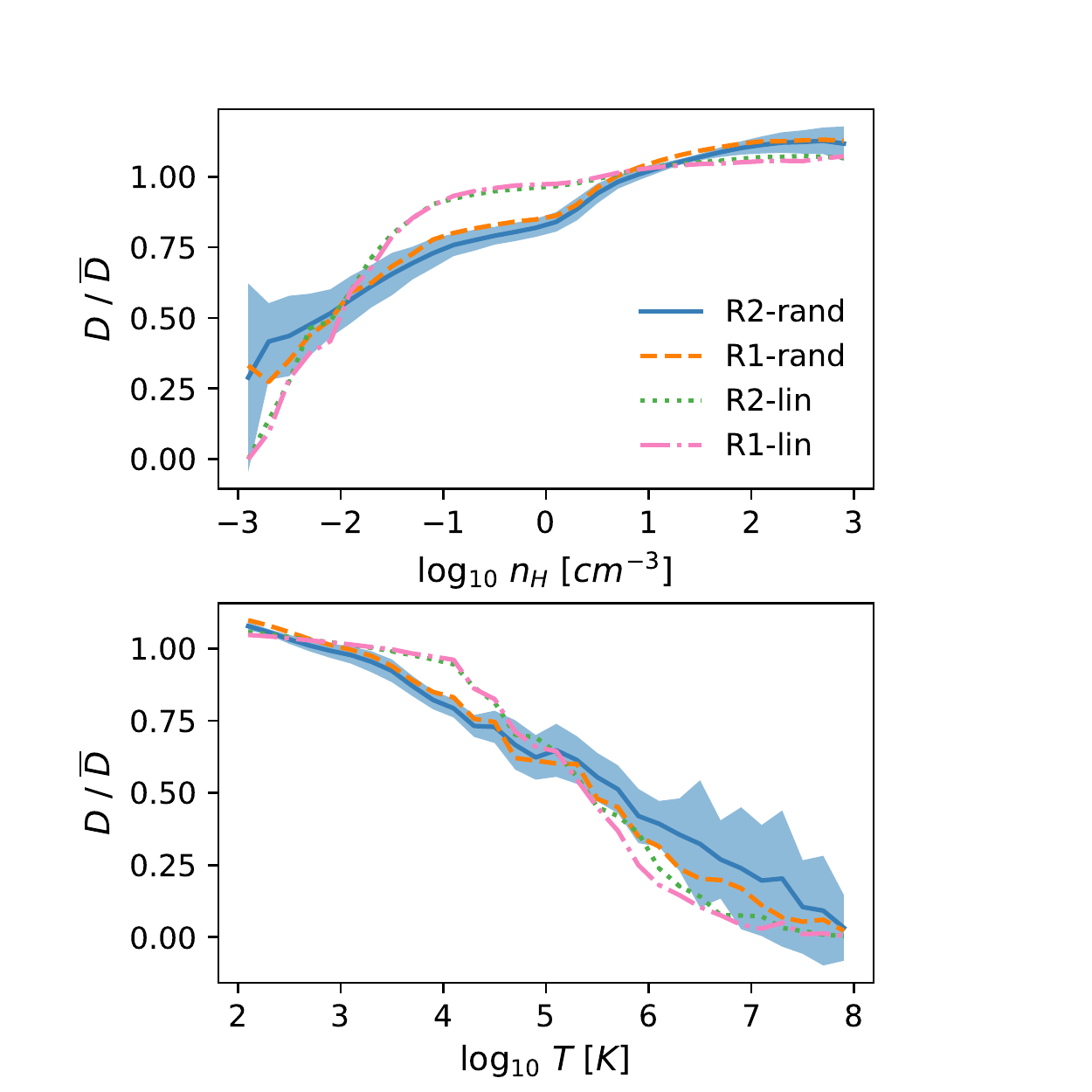}
	\caption{
		The local DGR normalized by the spatial DGR as a function of $n_{\rm H}$ (upper panel) and $T$ (lower panel), respectively,
		time-averaged from $t=0.1$Gyr to $t=0.4$Gyr.
		The temporal fluctuation ($\pm 1\sigma$) for the fiducial model is shown as the blue shaded region.
		The local DGR shows a positive correlation with $n_{\rm H}$ and an anti-correlation with $T$,
		as dust is destroyed in hot and diffuse gas.
	}
	\label{fig:rhoTvsDGR}
\end{figure}

In Fig. \ref{fig:rhoTvsDGR},
we show the local $D_{\rm}$ normalized by the spatial average $\overline{D}_{\rm}$  as a function of $n_{\rm H}$ (upper panel) and $T$ (lower panel), respectively, 
time-averaged from $t=0.1$ Gyr to 0.4 Gyr.
The temporal fluctuation ($\pm 1\sigma$) for the fiducial model is shown as the blue shaded region.
The local $D_{\rm}$ shows a positive correlation with $n_{\rm H}$ and an anti-correlation with $T$,
which is expected as dust is destroyed in hot and diffuse gas.
With linear driving,
$D_{\rm}$ is roughly homogeneous in most regions.
However,
since SNe occur predominantly in dense gas with linear driving,
dust destruction is very efficient (i.e. higher $\epsilon_{\rm sp}$),
leading to an abrupt decline of $D_{\rm}$ at $n_{\rm H} < 0.03 {\rm cm}^{-3}$ and $T > 10^4$K.
The SN bubbles have short dynamical times and will mix with the ambient medium rapidly.
In contrast,
with random driving,
SNe occur mostly in the diffuse gas and sometimes in the pre-existing bubbles,
which leads to less efficient dust destruction and slower gas mixing.
As a result,
the correlation exists even in the exterior of the bubbles as a consequence of incomplete gas mixing,
and there is a $\sim$ 30\% deficit of DGR in the volume filling warm gas ($n_{\rm H}\sim 0.1 {\rm cm^{-3}}$ and $T\sim 10^4$K) compared to that in the dense clouds.
The temporal fluctuation is largest in the hot and diffuse phase and decreases as density increases. It is about 8\% in the volume filling warm phase where $n_{\rm H}\sim 0.1 {\rm cm^{-3}}$.

\section{Discussion}\label{sec:discussion}

\subsection{Comparison with previous works}
In the following, we compare our results to previous studies of dust destruction in the ISM that mainly model dust sputtering in single SN shocks in a homogeneous ISM \citep[see][for a review]{Micelotta:2018gl}. Their estimates of $M_{\rm cl}$  can be directly compared to our single SN results described in Sect.3.2.2. The closest to our hydrodynamic simulations  is work by SDJ15 who considered dust sputtering in an evolving 1D SN blast wave. They find that $M_{\rm cl,C} = 1220 {\rm M_\odot}$ and $M_{\rm cl,Si} = 1990 {\rm M_\odot}$ for $n_{\rm H,0} = 0.25 {\rm cm^{-3}}$ when they assume $E_{51} = 1$, as in our model.
In our $n_{\rm H,0} = 0.3 {\rm cm^{-3}}$ case,
we find $M_{\rm cl,C} = 1180 {\rm M_\odot}$ and $M_{\rm cl,Si} = 1660 {\rm M_\odot}$,
which agrees very well with their results.
This is quite encouraging,
though it is important to note that there are a few differences between the two models.
For example,
they follow the evolution of the grain-size distribution while we assume a fixed one.
They also include a treatment for the magnetic pressure support which can suppress nonthermal sputtering.

Despite the good agreement on $M_{\rm cl}$,
SDJ15 estimate the dust destruction timescales to be $\tau_{\rm C} = 3.2$ Gyr and $\tau_{\rm Si} = 2$ Gyr,
which are significantly longer than ours ($\tau_{\rm C} = 0.44$ Gyr and $\tau_{\rm Si} = 0.35$ Gyr).
The discrepancy comes from several different assumptions about the ISM model rather than the dust destruction in individual SN.
In SDJ15,
the destruction timescale 
{follows the formulation proposed by \citet{1980ApJ...239..193D} as the time required to destroy all the dust in the ISM:}
\begin{equation}\label{eq:tSNest}
\tau = \frac{ M_{\rm gas} }{\delta_{\rm SN} R_{\rm SN}M_{\rm cl}},
\end{equation}
where $\delta_{\rm SN}$ is the SN correction factor that accounts for correlated SNe that explode within superbubbles and above the galactic disk and therefore do not destroy dust.
Their fiducial supernova energy is $E_{51} = 0.5$ which is only half of ours.
In this case, they obtain $M_{\rm cl,C}=600 {\rm M_\odot}$ and $M_{\rm cl,Si}=990 {\rm M_\odot}$, also about half of our values\footnote{In our simulations, SNe actually occur in a range of densities, mostly within $10^{-2} < n_{\rm H} < 1 {\rm cm^{-3}}$, rather than in a single density. However, $M_{\rm cl}$ is not very sensitive to $n_{\rm H}$ in this range, with the variation less than a factor of two.
}.
Another important difference is that their assumed SN rate per gas mass ($R_{\rm SN} / M_{\rm gas}$) is $1.4\times10^{-3} {\rm M_\odot}^{-1} {\rm Gyr}^{-1}$ (this includes their assumed volume filling factor of the warm gas which effectively reduces the SN rate by a factor of 0.8),
while our fiducial choice is $4\times10^{-3} {\rm M_\odot}^{-1} {\rm Gyr}^{-1}$ which is about 2.7 times higher.
The SN correction factor $\delta_{\rm SN}$ is 0.4 in our fiducial model (\textit{R2-rand}) which is very close to their 0.36 (which is taken from \citet{1989IAUS..135..431M}).
Therefore,
the main source of discrepancy comes from the adopted SN rate and the SN energy,
both of which are difficult to determine accurately from observations and thus dominate the uncertainty on $\tau$.


\cite{2011A&A...530A..44J} demonstrated that uncertainties in the observed  $M_{\rm gas}$ and $R_{\rm SN}$ cause the overall uncertainty in dust lifetime-estimates from Eq.~(\ref{eq:tSNest}) of the order of 90\%. For dust destruction results from \cite{1996ApJ...469..740J} for silicates and from \cite{SerraDiazCano:2008p588} for carbon grains, they yield values of  $0.03-1.0$~Gyr and  $0.02-0.51$~Gyr for $\tau_{\rm Si}$ and $\tau_{\rm C}$, respectively. The shorter lifetime of carbon dust compared to silicates is caused by their choice of hydrogenated amorphous carbon as carbonaceous material, which is less resilient than graphite adopted by \cite{1996ApJ...469..740J}. Our estimates for $\tau_{\rm C}$ and $\tau_{\rm Si}$ fall in the range of values derived by \cite{2011A&A...530A..44J}. The values of $M_{\rm cl,C}=1315 {\rm M_\odot}$ and $M_{\rm cl,Si}=1590 {\rm M_\odot}$ calculated for the destruction efficiencies from \cite{1996ApJ...469..740J} agree well with our results.

The destruction timescale for silicate grains has been recently evaluated with post-processing of 3D hydrodynamical simulations with dust evolution models including dust growth in the ISM  \citep{Zhukovska:2016jh}. They find that large variations of the Si abundances in the local Milky Way support destruction of silicate grains by SN shocks on a relatively short timescale of 0.35~Gyr, which is in excellent agreement with the value derived in the present work.

\subsection{Implications for sub-grid models and cosmological simulations}
The sub-grid models for dust destruction adopted in large-scale hydrodynamical simulations \citep{2013MNRAS.432.2298B, 2015MNRAS.449.1625B, 2017MNRAS.468.1505M, 2018MNRAS.478.2851M, 2018MNRAS.478.4905A} and semi-analytic models \citep{2017MNRAS.471.3152P} 
are mostly based on variations of Eq. \ref{eq:tSNest},
either used on a cell-by-cell basis or for an annulus within the galaxy.
In these simulations,
$R_{\rm SN}$ is no longer a free parameter but comes directly from the computed star formation rate,
and hence there are only two free parameters, $\delta_{\rm SN}$ and $M_{\rm cl}$, which are generally assumed to be constant.
Our results suggest that $\delta_{\rm SN}$ may not be constant and should vary with SN rate and SN environment.
In fact, $\delta_{\rm SN}$ can even be larger than unity if SNe occur in regions with higher than average DGR as suggested by
observations of the Magellanic Clouds \citep{2015ApJ...799..158T}. 
Incorporating dust destruction into simulations that can self-consistently follow star formation and the SN environment such as \citet{2017MNRAS.466.1903G, 2017ApJ...846..133K, 2019MNRAS.483.3363H} will be critical to systematically quantify the appropriate scaling for $\delta_{\rm SN}$.
In addition,
$M_{\rm cl}$ is expected to be variable due to its density dependence.
For example,
$M_{\rm cl}$ can be much smaller in a denser environment such as starburst galaxies or high-redshift galaxies.
Another complication is that $M_{\rm cl}$ is expected to depend on the grain-size distribution, which is not universal and may have evolved with cosmic time.

\citet{2017MNRAS.471.3152P} tracked dust production by AGB stars and SNe and via accretion in the ISM using the best available estimates of these rates, and dust destruction using the rates from SDJ15, in the context of a cosmological semi-analytic model of galaxy evolution. They found good agreement with observational estimates of dust masses at $z=0$, and found some tension with recent observations claiming that very large dust reservoirs were already in place at very early times ($z\sim 6$-7). Adopting the much shorter dust destruction times found in the work presented here will, on the face of it, greatly increase this tension and perhaps even cause difficulties in reproducing observational estimates of dust masses in nearby galaxies (see also the discussion in SDJ15, Section 4.4). 

\subsection{Missing physics and future improvements}

{As discussed in Sec. \ref{sec:size},
our model does not include dust production processes such as stellar ejecta and dust growth in the ISM,
which is important to follow the life cycle of the interstellar dust.
In addition,
processes that modify the size distribution but not the total dust mass such as shattering and coagulation need to be included to follow the evolution of the size distribution and understand its impact on dust destruction.
Another grain-grain collision process that takes place in SN shocks is evaporation,
though its contribution to dust destruction has generally been found to be subdominant (\citealp{1996ApJ...469..740J}).
}

Another potential improvement is to include the magnetic fields,
which should be straightforward as the {\sc Gizmo} code is able to solve magnetohydrodynamics (MHD) and the magnetic fields can be evolved self-consistently.
The assumption of $B\propto \rho_{\rm gas}$ in shocks can then be relaxed and the information on $B$ will be directly available when calculating the betatron acceleration.
{In addition, the magnetic pressure provides extra support against the shell compression and therefore reduces nonthermal sputtering via betatron acceleration, though the exact behavior will depend on the configuration of the magnetic fields.}
It is also possible to directly integrate the gyration of grains (at least for the large grains) which changes the direction of ${\bf v_{rel}}$ but not its magnitude,
which may have some effect on nonthermal sputtering in interacting shocks.

{Our grain charge model is very simplistic.
A more sophisticated estimate such as \citet{1987ApJ...318..674M} or the more recent work of \citet{2001ApJS..134..263W} can be adopted to improve the calculation of the plasma drag,}
though we do not expect the results to change significantly as plasma drag is sub-dominant in all but the most diffuse cases ($n_{\rm H,0} \lesssim 0.01 {\rm cm^{-3}}$),
{where it is not only rare for SNe to occur (probability $< 10\%$) but also sputtering is inefficient ($\epsilon_{\rm sp} < 0.05$).
}

{Finally,
as discussed in Sec. \ref{sec:sputter},
the distinction between thermal and nonthermal sputtering is artificial and can be replaced with a more natural formulation based on the skewed Maxwellian distribution.
Future studies along the line of \citet{2014A&A...570A..32B} will be very valuable if a fine-grid table of the calculated erosion rate as a function of both the drift velocity of dust and the gas temperature can be provided.
}

\section{Summary}\label{sec:summary}

We have introduced a novel numerical framework to follow dust sputtering in hydrodynamical simulations in a more \textit{ab initio} fashion.
We adopt a one-fluid approach where dust is spatially coupled with the gas, which is justified due to the small Larmor radius in the ISM.
In order to follow nonthermal sputtering,
we solve the equation of motion for dust relative to the gas which is controlled by direct collisions, plasma drag and betatron acceleration.
We use a subcycling technique to tackle the stiffness problem for integrating the dust mass and the dust-gas relative velocity, 
which can be bypassed when the sputtering rate becomes low.
We adopt an MRN grain-size distribution and do the integration bin-by-bin assuming that the distribution is remains constant in time.

We have systematically investigated dust destruction for a single SN occurring in an initially uniform medium.
Dust destruction is primarily due to thermal sputtering for small grains and nonthermal sputtering for large grains (Fig. \ref{fig:fig2_Msp_vs_time} and \ref{fig:fig3finalmsp}).
The grain-size integrated sputtering is dominated by nonthermal sputtering in the range of $n_{\rm H,0}$ where SNe typically occur (Fig. \ref{fig:fig4finalmspvsnh}).
We provide fitting formulae for $M_{\rm cl}$ as a function of $n_{\rm H,0}$ (Eq. \ref{eq:fitting}).
The ratio of $M_{\rm cl,Si}$ to $M_{\rm cl,C}$ is around 1.5 at low densities and it decreases as $n_{\rm H,0}$ increases
because sputtering becomes limited by the available dust in the shocked gas.

We have conducted the first hydrodynamical simulations that explicitly follow dust sputtering in a turbulent multiphase ISM (Fig. \ref{fig:dgrmap}).
The dust destruction timescales in the simulations are $\tau_{\rm C} = 0.44$ Gyr and $\tau_{\rm Si} = 0.35$ Gyr for our fiducial model and they scale sub-linearly with $R_{\rm SN}^{-1}$.
SNe that occur in the pre-existing low-DGR bubbles destroy less dust as sputtering becomes limited by the available dust in the bubbles (Fig. \ref{fig:tSN_estimate}).
This effectively increases the destruction timescales by a factor of $\delta_{\rm SN}^{-1} \sim 2.5$ compared to estimates based on our single SN results (Fig. \ref{fig:applysinglesn}).
Sputtering leads to a spatial inhomogeneity of DGR $z_{D}\sim$14\% for scales below 10 pc (Fig. \ref{fig:stdvspixelsize}).
Locally, the DGR correlates positively with gas density and negatively with gas temperature even in the exterior of the bubbles as a consequence of incomplete gas mixing,
leading to a $\sim$ 30\% DGR deficit in the volume filling warm gas compared to that in the dense clouds (Fig. \ref{fig:rhoTvsDGR}).


\section*{Acknowledgments}
We thank the referee, Jonathan Slavin, for his insightful comments which helped improve our paper.
We further thank Eli Dwek, Chris McKee and Hiroyuki Hirashita for valuable discussions.
as well as Volker Springel, Phil Hopkins and the Grackle team for making {\sc Gadget-3}, {\sc Gizmo} and {\sc Grackle} codes publicly available.
We use {\sc pygad}\footnote{https://bitbucket.org/broett/pygad} for visualization.
The Center for Computational Astrophysics is supported by the Simons Foundation.

\bibliographystyle{mn2e}
\bibliography{literatur}

\appendix
\section{Convergence test} \label{app:convtest}

\begin{figure*}
	\centering
	\includegraphics[trim = 0mm 5mm 10mm 0mm, clip,width=0.9\linewidth]{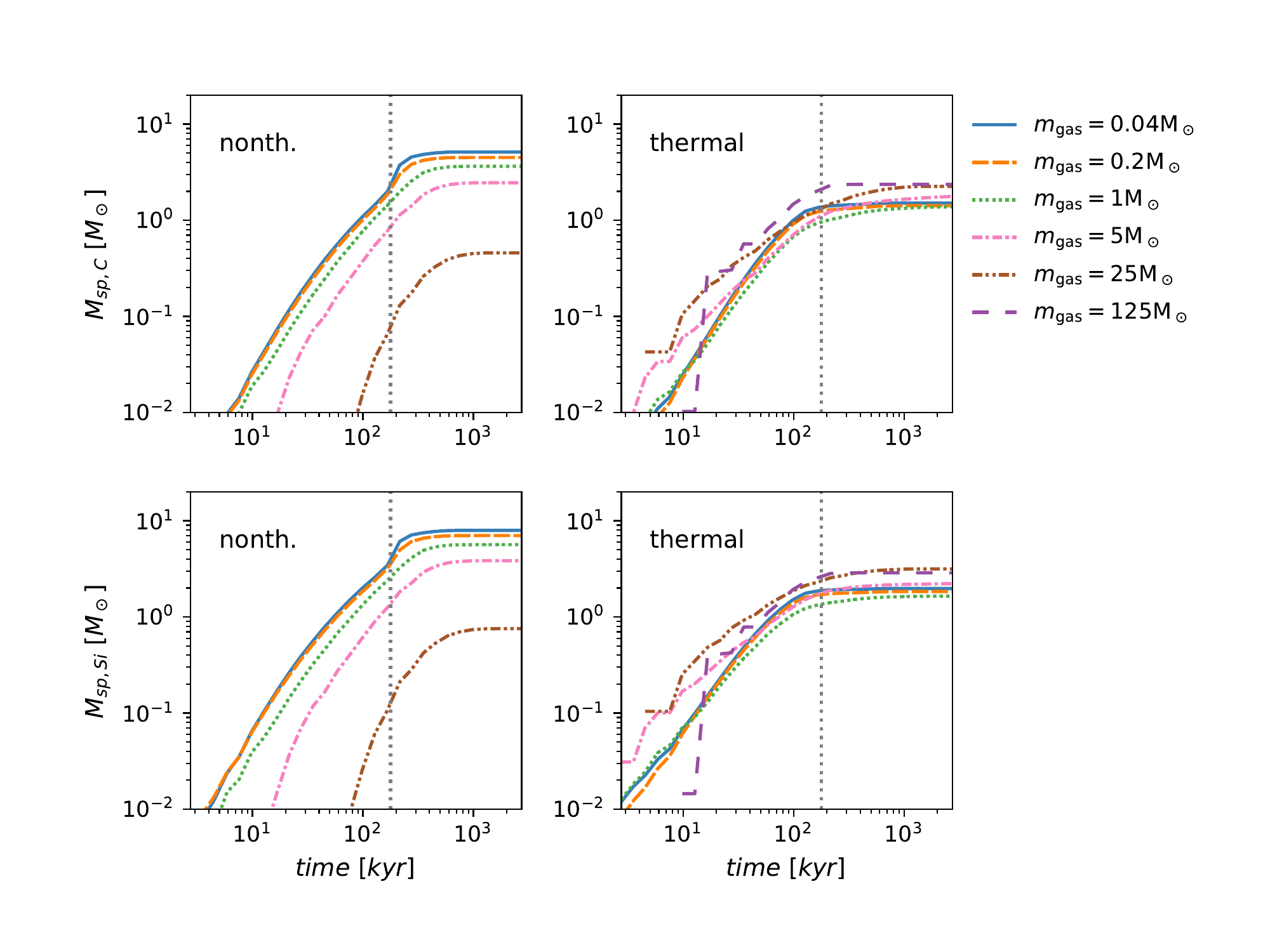}
	\caption{
		Convergence test for a single SN for $n_{\rm H,0} = 0.1{\rm cm^{-3}}$.
		Thermal sputtering converges at $m_{\rm gas} = 5 {\rm M_\odot}$,
		while nonthermal sputtering converges at $m_{\rm gas} = 0.2 {\rm M_\odot}$,
		as the latter is controlled by ${\rm v_{rel}}$
		which depends sensitively on the shock structure.
		Compared to the dynamical impact of SNe which only requires resolving the cooling radius,
		sputtering is much more difficult to resolve.
	}
	\label{fig:conv}
\end{figure*}

We conduct a convergence test for a single SN with $n_{\rm H,0} = 0.1 {\rm cm^{-3}}$.
In Fig. \ref{fig:conv},
we show the time evolution of $M^{\rm nth}_{\rm sp,C}$ (upper left), 
$M^{\rm th}_{\rm sp,C}$ (upper right),
$M^{\rm nth}_{\rm sp,Si}$ (lower left)
and $M^{\rm nth}_{\rm sp,Si}$ (lower right), respectively.
The gas particle mass is systematically increased from 0.04 M$_\odot$ to 125 M$_\odot$.
A thermal energy of $10^{51}$ erg is injected into the $N_{\rm inj} = \max(32, 16{\rm M_\odot}/m_{\rm gas})$ nearest gas particles in a kernel-weighted fashion.

Thermal sputtering converges at $m_{\rm gas} = 5 {\rm M_\odot}$,
and coarsening the resolution further up to $m_{\rm gas} = 125 {\rm M_\odot}$ only
overestimates the sputtered mass by a factor of two.
This is because thermal sputtering is controlled by the temperature. 
As long as the initial injection of SN is able to heat the gas temperature up to $T\gtrsim 10^6$K where thermal sputtering is efficient,
the final sputtered mass will not be very different.
For $m_{\rm gas} = 125 {\rm M_\odot}$,
the sputtered mass is even higher as the shocked gas mass is overestimated during the injection of SN energy because of the poor resolution.
Once the resolution becomes even coarser such that no gas will be heated above $T>10^6$K,
the sputtered mass will abruptly drop to essentially zero.
On the other hand,
convergence for nonthermal sputtering is much more computationally demanding:
it becomes increasingly efficient as $m_{\rm gas}$ decreases and eventually shows convergence at $m_{\rm gas} = 0.2 {\rm M_\odot}$.
This is because nonthermal sputtering is controlled by ${\rm v_{rel}}$
which depends sensitively on the shock structure.
Coarsening the resolution smooths out the shock structure and therefore suppresses nonthermal sputtering.
Since nonthermal sputtering dominates over thermal sputtering in most cases,
the convergence criterion for total sputtering is also $m_{\rm gas} = 0.2 {\rm M_\odot}$.
Compared to the dynamical impact of SNe which only requires resolving the cooling radius,
sputtering is much more difficult to resolve.

\section{Analytic solutions for dust-gas relative velocity} \label{app:analyticvel}
For completeness, we give the analytic solutions of ${\rm v_{rel}}$ that we adopted to validate our time integration schemes (cf. Fig. \ref{fig:dragtest}).
For the linear drag relation with a constant $t_{\rm drag}$ and $a_{\rm hydro}$,
the equation of motion is
\begin{eqnarray}
\frac{\rm d v_{rel}}{{\rm d}t} = 	- \frac{{\rm v_{rel}} }{t_{\rm drag}} - {a_{\rm hydro}},
\end{eqnarray}
which has an analytic solution:
\begin{eqnarray}\label{eq:linearV}
	{\rm v_{rel}} (t) = {\rm v_{rel}}(0) e^{-t / t_{\rm drag}} - a_{\rm hydro} t_{\rm drag} ( 1 - e^{-t / t_{\rm drag}} ).
\end{eqnarray}
As $t\to \infty$, the solution approaches the terminal velocity ${\rm v_{rel}} = -a_{\rm hydro} t_{\rm drag}$.

In the supersonic regime, the equation of motion becomes nonlinear due to the quadratic term:
\begin{eqnarray}
	\frac{\rm d v_{rel}}{{\rm d}t} = 	- \frac{{\rm v_{rel}} \sqrt{1+C {\rm v^2_{rel}}}}{t_{\rm drag}} - {a_{\rm hydro}},
\end{eqnarray}
where $C = 0.22c_s^{-2}$. 
An analytic solution is not available in this case. 
However, if we assume $a_{\rm hydro} = 0$,
an analytic relation between ${\rm v_{rel}}$ and $t$ can be expressed as
\begin{eqnarray}\label{eq:nonlinearV}
	t = 0.5 t_{\rm drag} \Big[ \ln|C {\rm v^2_{rel}}(0) - 1| - \ln(C {\rm v^2_{rel}}(0) + 1) \nonumber\\
											- \ln|C {\rm v^2_{rel}}(t) - 1| + \ln(C {\rm v^2_{rel}}(t) + 1) \Big].
\end{eqnarray}

\end{document}